\documentclass[twocolumn]{aastex62}

\usepackage{longtable}
\usepackage{graphicx}
\usepackage{amsmath,amssymb}
\usepackage{color}
\usepackage{units}
\usepackage{epstopdf}
\usepackage{hyperref}
\usepackage{multirow}
\usepackage{url}

\usepackage{subfigure}
\usepackage{rotating}

\newcommand{\beq}{\begin{equation}}
\newcommand{\eeq}{\end{equation}}
\newcommand{\bdm}{\begin{displaymath}}
\newcommand{\edm}{\end{displaymath}}

\definecolor{Gray}{gray}{0.9}
\definecolor{orange}{rgb}{0.9,0.5,0}

\graphicspath{{./plots/}}

\begin{document}

\title{GROWTH on S190425z: Searching thousands of square degrees to identify an optical or infrared counterpart to a binary neutron star merger with the Zwicky Transient Facility and Palomar Gattini IR}

\author[0000-0002-8262-2924]{Michael W. Coughlin}
\affil{Division of Physics, Mathematics, and Astronomy, California Institute of Technology, Pasadena, CA 91125, USA}

\author[0000-0002-2184-6430]{Tom{\'a}s Ahumada}
\affil{Department of Astronomy, University of Maryland, College Park, MD 20742, USA}

\author{Shreya Anand}
\affil{Division of Physics, Mathematics, and Astronomy, California Institute of Technology, Pasadena, CA 91125, USA}

\author{Kishalay De}
\affil{Division of Physics, Mathematics, and Astronomy, California Institute of Technology, Pasadena, CA 91125, USA}

\author[0000-0001-9315-8437]{Matthew J. Hankins}
\affiliation{Division of Physics, Mathematics, and Astronomy, California Institute of Technology, Pasadena, CA 91125, USA}

\author{Mansi M. Kasliwal}
\affil{Division of Physics, Mathematics, and Astronomy, California Institute of Technology, Pasadena, CA 91125, USA}

\author[0000-0001-9898-5597]{Leo P. Singer}
\affiliation{Astrophysics Science Division, NASA Goddard Space Flight Center, MC 661, Greenbelt, MD 20771, USA}
\affiliation{Joint Space-Science Institute, University of Maryland, College Park, MD 20742, USA}

\author[0000-0001-8018-5348]{Eric C. Bellm}
\affiliation{DIRAC Institute, Department of Astronomy, University of Washington, 3910 15th Avenue NE, Seattle, WA 98195, USA}

\author[0000-0002-8977-1498]{Igor~Andreoni}
\affil{Division of Physics, Mathematics, and Astronomy, California Institute of Technology, Pasadena, CA 91125, USA}

\author[0000-0003-1673-970X]{S. Bradley Cenko}
\affiliation{Astrophysics Science Division, NASA Goddard Space Flight Center, MC 661, Greenbelt, MD 20771, USA}
\affiliation{Joint Space-Science Institute, University of Maryland, College Park, MD 20742, USA}

\author[0000-0001-5703-2108]{Jeff Cooke}
\affiliation{Australian Research Council Centre of Excellence for Gravitational Wave Discovery (OzGrav), Swinburne University of Technology, Hawthorn, VIC, 3122, Australia}
\affiliation{Centre for Astrophysics and Supercomputing, Swinburne University of Technology, Hawthorn, VIC, 3122, Australia}

\author[0000-0001-7983-8698]{Christopher M. Copperwheat}
\affil{Astrophysics Research Institute, Liverpool John Moores University, \\ IC2, Liverpool Science Park, 146 Brownlow Hill, Liverpool L3 5RF, UK}

\author{Alison M. Dugas}
\affiliation{Division of Physics, Mathematics, and Astronomy, California Institute of Technology, Pasadena, CA 91125, USA}

\author[0000-0001-5754-4007]{Jacob E.\ Jencson}
\affiliation{Division of Physics, Mathematics, and Astronomy, California Institute of Technology, Pasadena, CA 91125, USA}

\author{Daniel A. Perley}
\affiliation{Astrophysics Research Institute, Liverpool John Moores University, \\ IC2, Liverpool Science Park, 146 Brownlow Hill, Liverpool L3 5RF, UK}


\author[0000-0001-8894-0854]{Po-Chieh Yu}
\affiliation{Graduate Institute of Astronomy, National Central University, 32001, Taiwan}

\author[0000-0002-6112-7609]{Varun Bhalerao}
\affiliation{Indian Institute of Technology Bombay, Powai, Mumbai 400076, India}

\author{Harsh Kumar}
\affiliation{Indian Institute of Technology Bombay, Powai, Mumbai 400076, India}

\author[0000-0002-7777-216X]{Joshua S. Bloom}
\affiliation{Department of Astronomy, University of California, Berkeley, CA 94720-3411, USA; Physics, Lawrence Berkeley National Laboratory, 1 Cyclotron Road, MS 50B-4206, Berkeley, CA 94720, USA}


\author{G.C. Anupama}
\affiliation{Indian Institute of Astrophysics, II Block Koramangala, Bengaluru 560034, India}

\author{Michael C. B. Ashley}
\affil{School of Physics, University of New South Wales, Sydney NSW 2052, Australia}

\author{Ashot Bagdasaryan}
\affil{Division of Physics, Mathematics, and Astronomy, California Institute of Technology, Pasadena, CA 91125, USA}

\author[0000-0002-5741-7195]{Rahul Biswas}
\affiliation{The Oskar Klein Centre, Department of Physics, Stockholm University, AlbaNova, SE-106 91 Stockholm, Sweden}

\author{David A.\ H.\ Buckley}
\affiliation{South African Astronomical Observatory, P.O. Box 9, Observatory 7935, Cape Town, South Africa}
\affiliation{Southern African Large Telescope Foundation, P.O. Box 9, Observatory 7935, Cape Town, South Africa}

\author[0000-0002-7226-836X]{Kevin B. Burdge}
\affil{Division of Physics, Mathematics, and Astronomy, California Institute of Technology, Pasadena, CA 91125, USA}

\author[0000-0002-6877-7655]{David O. Cook}
\affiliation{IPAC, California Institute of Technology, 1200 E. California Blvd, Pasadena, CA 91125, USA}

\author{John Cromer}
\affiliation{Caltech Optical Observatories, California Institute of Technology, Pasadena, CA 91125, USA}

\author[0000-0003-2292-0441]{Virginia Cunningham}
\affiliation{Department of Astronomy, University of Maryland, College Park, MD 20742, USA}

\author{Antonino D'A\`i}
\affiliation{INAF/IASF-Palermo, via Ugo La Malfa 153, I-90146, Palermo, Italy}

\author{Richard G. Dekany}
\affiliation{Caltech Optical Observatories, California Institute of Technology, Pasadena, CA 91125, USA}

\author{Alexandre Delacroix}
\affiliation{Caltech Optical Observatories, California Institute of Technology, Pasadena, CA 91125, USA}

\author{Simone Dichiara}
\affiliation{Astrophysics Science Division, NASA Goddard Space Flight Center, MC 661, Greenbelt, MD 20771, USA}
\affiliation{Department of Astronomy, University of Maryland, College Park, MD 20742, USA}

\author[0000-0001-5060-8733]{Dmitry A. Duev}
\affiliation{Division of Physics, Mathematics, and Astronomy, California Institute of Technology, Pasadena, CA 91125, USA}

\author{Anirban Dutta}
\affiliation{Indian Institute of Astrophysics, II Block Koramangala, Bengaluru 560034, India}

\author{Michael Feeney}
\affiliation{Caltech Optical Observatories, California Institute of Technology, Pasadena, CA 91125, USA}

\author[0000-0001-9676-730X]{Sara Frederick}
\affiliation{Department of Astronomy, University of Maryland, College Park, MD 20742, USA}

\author[0000-0002-1955-2230]{Pradip Gatkine}
\affiliation{Department of Astronomy, University of Maryland, College Park, MD 20742, USA}

\author{Shaon Ghosh}
\affiliation{Center for Gravitation, Cosmology and Astrophysics, Department of Physics, University of Wisconsin--Milwaukee, P.O.\ Box 413, Milwaukee, WI 53201, USA}

\author[0000-0003-3461-8661]{Daniel~A.~Goldstein}
\affiliation{Division of Physics, Mathematics, and Astronomy, California Institute of Technology, Pasadena, CA 91125, USA}

\author[0000-0001-8205-2506]{V. Zach Golkhou}
\altaffiliation{Moore-Sloan, WRF, and DIRAC Fellow}
\affiliation{DIRAC Institute, Department of Astronomy, University of Washington, 3910 15th Avenue NE, Seattle, WA 98195, USA}
\affiliation{The eScience Institute, University of Washington, Seattle, WA 98195, USA}

\author[0000-0002-4163-4996]{Ariel Goobar}
\affiliation{The Oskar Klein Centre, Department of Physics, Stockholm University, AlbaNova, SE-106 91 Stockholm, Sweden}

\author[0000-0002-3168-0139]{Matthew J. Graham}
\affiliation{Division of Physics, Mathematics, and Astronomy, California Institute of Technology, Pasadena, CA 91125, USA}

\author[0000-0001-8221-6048]{Hidekazu Hanayama}
\affiliation{Ishigakijima Astronomical Observatory, National Astronomical Observatory of Japan, 1024-1 Arakawa, Ishigaki, Okinawa 907-0024, Japan}

\author{Takashi Horiuchi}
\affiliation{Ishigakijima Astronomical Observatory, National Astronomical Observatory of Japan, 1024-1 Arakawa, Ishigaki, Okinawa 907-0024, Japan}

\author{Tiara Hung}
\affiliation{Department of Astronomy and Astrophysics, University of California, Santa Cruz, CA 95064, USA}

\author[0000-0001-8738-6011]{Saurabh W. Jha}
\affiliation{Department of Physics and Astronomy, Rutgers, the State University of New Jersey,136 Frelinghuysen Rd., Piscataway, NJ 08854, USA}
\affiliation{Center for Computational Astrophysics, Flatiron Institute, 162 5th Avenue, New York, NY 10010, USA}

\author[0000-0002-5105-344X]{Albert K. H. Kong}
\affiliation{Institute of Astronomy, National Tsing Hua University, Hsinchu 30013, Taiwan}

\author{Matteo Giomi}
\affiliation{Humboldt Universitaet zu Berlin, Newtonstra{\ss}e 15, 12489 Berlin, Germany}

\author[0000-0001-6295-2881]{David~L.\ Kaplan}
\affiliation{Center for Gravitation, Cosmology and Astrophysics, Department of Physics, University of Wisconsin--Milwaukee, P.O.\ Box 413, Milwaukee, WI 53201, USA}

\author{V. R. Karambelkar }
\affiliation{Indian Institute of Technology Bombay, Powai, Mumbai 400076, India}

\author{Marek Kowalski}
\affiliation{Institute of Physics, Humboldt-Universit\"at zu Berlin, Newtonstr. 15, 124 89 Berlin, Germany}
\affiliation{Deutsches Elektronensynchrotron, Platanenallee 6, D-15738, Zeuthen, Germany}

\author[0000-0001-5390-8563]{Shrinivas R. Kulkarni}
\affiliation{Division of Physics, Mathematics, and Astronomy, California Institute of Technology, Pasadena, CA 91125, USA}

\author{Thomas Kupfer}
\affiliation{Kavli Institute for Theoretical Physics, University of California, Santa Barbara, CA 93106, USA}

\author{Frank J. Masci}
\affiliation{IPAC, California Institute of Technology, 1200 E. California
             Blvd, Pasadena, CA 91125, USA}

\author{Paolo Mazzali}
\affiliation{Astrophysics Research Institute, Liverpool John Moores University, \\ IC2, Liverpool Science Park, 146 Brownlow Hill, Liverpool L3 5RF, UK}

\author{Anna M. Moore}
\affiliation{Research School of Astronomy and Astrophysics, Australian National University, Canberra, ACT 2611, Australia}

\author{Moses Mogotsi}
\affiliation{Southern African Large Telescope Foundation, P.O. Box 9, Observatory 7935, Cape Town, South Africa}
\affiliation{South African Astronomical Observatory, P.O. Box 9, Observatory 7935, Cape Town, South Africa}

\author{James D. Neill}
\affil{Division of Physics, Mathematics, and Astronomy, California Institute of Technology, Pasadena, CA 91125, USA}

\author[0000-0001-8771-7554]{Chow-Choong Ngeow}
\affiliation{Graduate Institute of Astronomy, National Central University, 32001, Taiwan}

\author{Jorge Mart\'inez-Palomera}
\affiliation{Department of Astronomy, University of California, Berkeley, CA 94720-3411, USA}

\author{Valentina La Parola}
\affiliation{INAF/IASF-Palermo, via Ugo La Malfa 153, I-90146, Palermo, Italy}

\author{M. Pavana}
\affiliation{Indian Institute of Astrophysics, II Block Koramangala, Bengaluru 560034, India}

\author{Eran O. Ofek}
\affiliation{Department of Particle Physics \& Astrophysics, Weizmann Institute of Science,
Rehovot 76100, Israel}

\author{Atharva Sunil Patil}
\affiliation{Graduate Institute of Astronomy, National Central University, 32001, Taiwan}

\author{Reed Riddle}
\affiliation{Caltech Optical Observatories, California Institute of Technology, Pasadena, CA 91125, USA}        

\author{Mickael Rigault}
\affiliation{Universit\'e Clermont Auvergne, CNRS/IN2P3, Laboratoire de Physique de Clermont, F-63000 Clermont-Ferrand, France}

\author{Ben Rusholme}
\affiliation{IPAC, California Institute of Technology, 1200 E. California
             Blvd, Pasadena, CA 91125, USA}

\author{Eugene Serabyn}
\affiliation{Jet Propulsion Laboratory, California Institute of Technology, Pasadena, CA 91109, USA}

\author[0000-0003-4401-0430]{David L. Shupe}
\affiliation{IPAC, California Institute of Technology, 1200 E. California
             Blvd, Pasadena, CA 91125, USA}

\author{Yashvi Sharma }
\affiliation{Indian Institute of Technology Bombay, Powai, Mumbai 400076, India}

\author{Avinash Singh}
\affiliation{Indian Institute of Astrophysics, II Block Koramangala, Bengaluru 560034, India}

\author{Jesper Sollerman}
\affiliation{The Oskar Klein Centre \& Department of Astronomy, Stockholm University, AlbaNova, SE-106 91 Stockholm, Sweden}

\author{Jamie Soon}
\affiliation{Research School of Astronomy and Astrophysics, Australian National University, Canberra, ACT 2611, Australia}

\author{Kai Staats}
\affiliation{Center for Interdisciplinary Exploration and Research in Astrophysics and Department of Physics and Astronomy, Northwestern University, 2145 Sheridan Road, Evanston, IL 60208, USA}

\author{Kirsty Taggart}
\affiliation{Astrophysics Research Institute, Liverpool John Moores University, \\ IC2, Liverpool Science Park, 146 Brownlow Hill, Liverpool L3 5RF, UK}

\author{Hanjie Tan}
\affiliation{Graduate Institute of Astronomy, National Central University, 32001, Taiwan}

\author{Tony Travouillon}
\affiliation{Research School of Astronomy and Astrophysics, Australian National University, Canberra, ACT 2611, Australia}

\author{Eleonora Troja}
\affiliation{Department of Astronomy, University of Maryland, College Park, MD 20742, USA}
\affiliation{Astrophysics Science Division, NASA Goddard Space Flight Center, MC 661, Greenbelt, MD 20771, USA}

\author{Gaurav Waratkar}
\affiliation{Indian Institute of Technology Bombay, Powai, Mumbai 400076, India}

\author[0000-0003-1890-3913]{Yoichi Yatsu}
\affiliation{Department of Physics, Tokyo Institute of Technology, 2-12-1, Ookayama, Meguro, Tokyo 152-8551, Japan}

\begin{abstract}
The third observing run by LVC has brought the discovery of many compact binary coalescences. Following the detection of the first binary neutron star merger in this run (LIGO/Virgo S190425z), we performed a dedicated follow-up campaign with the Zwicky Transient Facility (ZTF) and Palomar Gattini-IR telescopes. The initial skymap of this single-detector gravitational wave (GW) trigger spanned most of the sky observable from Palomar Observatory. Covering 8000 deg$^2$ of the initial skymap over the next two nights, corresponding to 46\% integrated probability, ZTF system achieved a depth of $\approx$\,21 $m_\textrm{AB}$ in $g$- and $r$-bands. Palomar Gattini-IR covered 2200 square degrees in $J$-band to a depth of 15.5\,mag, including 32\% integrated probability based on the initial sky map. The revised skymap issued the following day reduced these numbers to 21\% for the Zwicky Transient Facility and 19\% for Palomar Gattini-IR.
We narrowed 338,646 ZTF transient ``alerts'' over the first two nights of observations to 15 candidate counterparts.
Two candidates, ZTF19aarykkb and ZTF19aarzaod, were particularly compelling given that their location, distance, and age were consistent with the GW event, and their early optical lightcurves were photometrically consistent with that of kilonovae. These two candidates were spectroscopically classified as young core-collapse supernovae. The remaining candidates were ruled-out as supernovae. Palomar Gattini-IR did not identify any viable candidates with multiple detections only after merger time. We demonstrate that even with single-detector GW events localized to thousands of square degrees, systematic kilonova discovery is feasible.

\end{abstract}


\section{Introduction}

\begin{figure*}[t]
 \includegraphics[width=3.5in]{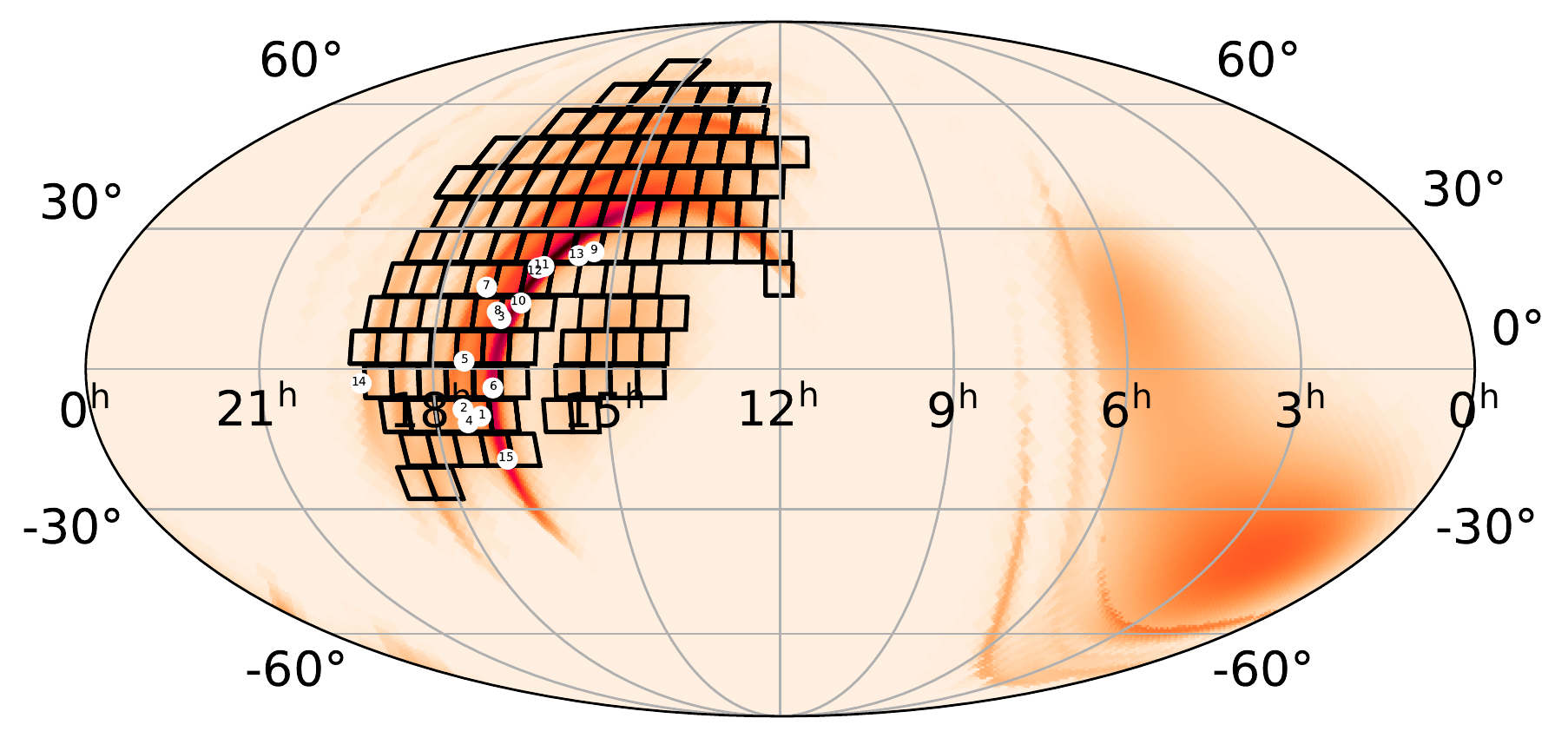}
 \includegraphics[width=3.5in]{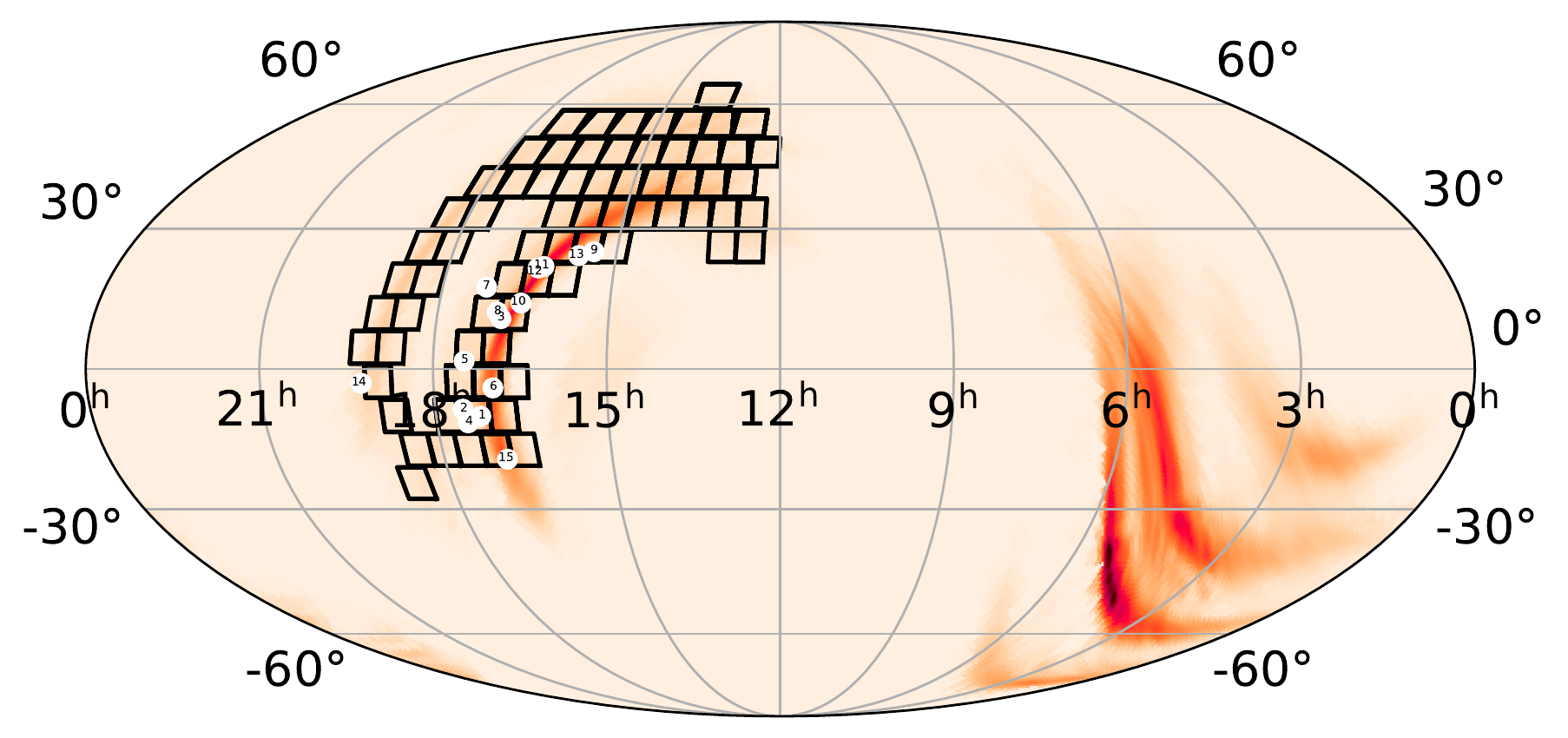} 
 \includegraphics[width=3.5in]{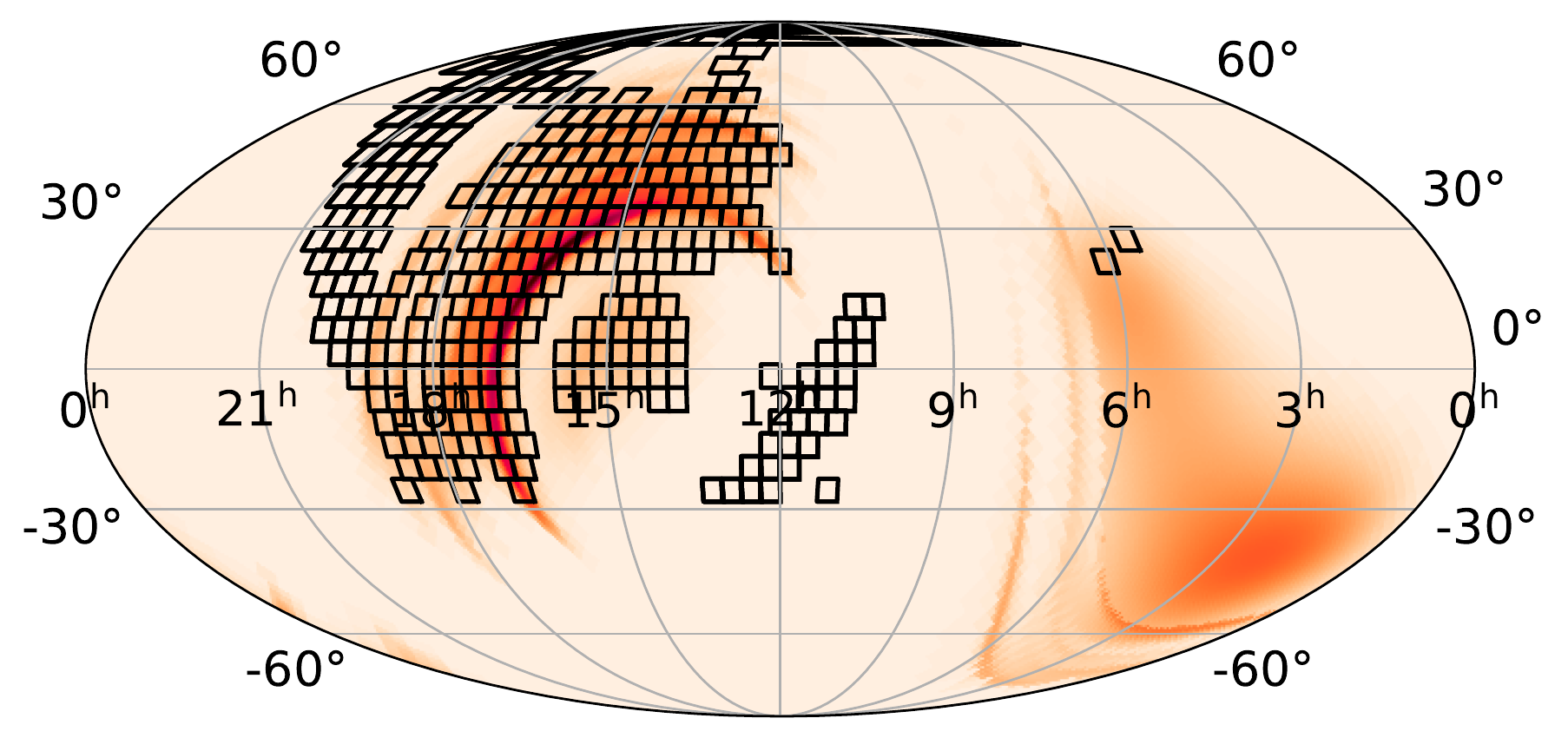}
 \includegraphics[width=3.5in]{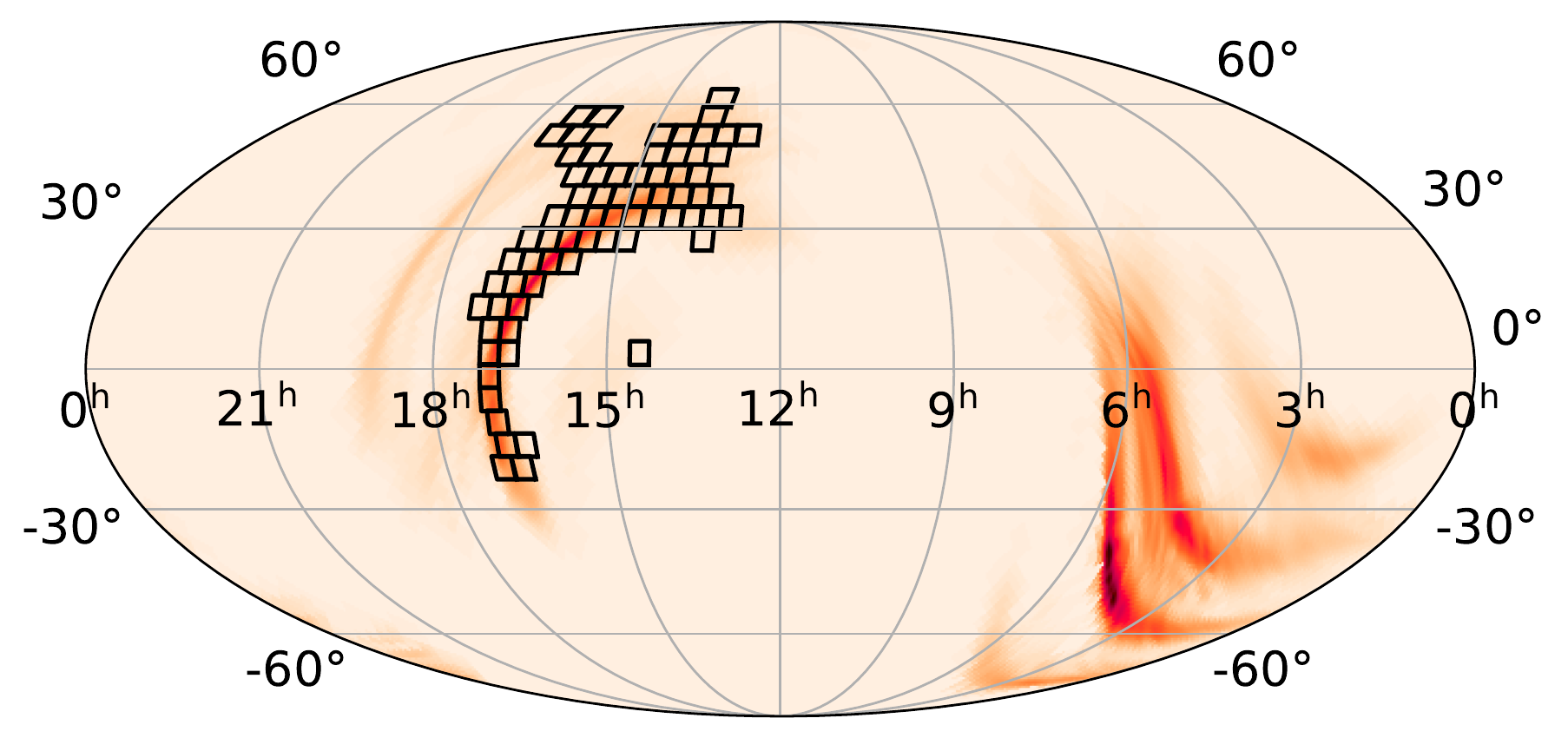} 
  \caption{Coverage of S190425z. (Left) The top and bottom rows show the $\approx$ 47 deg$^2$ ZTF tiles and the $\approx$ 25 deg$^2$ Palomar Gattini-IR tiles respectively on the 90\% probability region of the initial BAYESTAR skymap, along with the identified transients highlighted in Table~\ref{table:followup}. For the ZTF observations, the numbering scheme is 1: ZTF19aarykkb, 2: ZTF19aarzaod, 3: ZTF19aasckwd, 4: ZTF19aasfogv, 5: ZTF19aasejil, 6: ZTF19aaryxjf, 7: ZTF19aascxux, 8: ZTF19aasdajo, 9: ZTF19aasbamy, 10: ZTF19aasckkq, 11: ZTF19aarycuy, 12: ZTF19aasbphu, 13: ZTF19aasbaui, 14: ZTF19aarxxwb, 15: ZTF19aashlts.  (Right) We show the tilings of the two telescopes on the final LALInference map. We only include the tiles in the inner 90\% probability region for each skymap.}
 \label{fig:skymap}
\end{figure*} 
The third observing run (O3) by the network of gravitational-wave (GW) detectors with Advanced LIGO \citep{aLIGO} and Advanced Virgo \citep{adVirgo} began in April 2019.  This detector network has already observed over a score binary black holes thus far \citep{SiEA2019,ShEA2019,ChEA2019,SiEA2019a,ChEA2019a,GhEA2019}.
The current discovery rate builds on the success of the first few observing runs, which yielded 10 binary black hole detections \citep{AbEA2018b}. 

In addition, the coincident discovery of the binary neutron star (BNS) merger GW170817~\citep{AbEA2017b}, 
a short gamma-ray burst (SGRB) GRB170817A~\citep{AbEA2017e,GoVe2017,SaFe2017}, 
with an afterglow \citep{2017ApJ...848L..21A,2017ApJ...848L..25H,2017Sci...358.1579H,2017ApJ...848L..20M,2017Natur.551...71T} and ``kilonova'' (KN) counterpart, AT2017gfo \citep{ChBe2017,2017Sci...358.1556C,CoBe2017,2017Sci...358.1570D,2017Sci...358.1565E,KaNa2017,KiFo2017,LiGo2017,2017ApJ...848L..32M,NiBe2017,2017Sci...358.1574S,2017Natur.551...67P,SmCh2017,2017PASJ...69..101U}, initiated a new era of multi-messenger astronomy. Amongst many other science cases, measurements of the equation of state (EOS) of neutron stars \citep{BaBa2013,AbEA2017b,RaPe2018,BaJu2017,CoDi2018b}, 
the formation of heavy elements \citep{JuBa2015,WuFe2016,RoLi2017,AbEA2017f,RoFe2017,KaKa2019},
and the expansion rate of the universe \citep{2017Natur.551...85A,HoNa2018,CoDi2019} are all important results of the first BNS detection.

Following the success of GW170817, the Zwicky Transient Facility (ZTF) \citep{Bellm2018,Graham2018,DeSm2018,MaLa2018} on the Palomar 48 inch telescope, and Palomar Gattini-IR, a new wide-field near-infrared survey telescope at Palomar observatory, have been observing both SGRBs from the {\it Fermi} Gamma-ray Burst Monitor \citep{CoEA2018a,CeEA2018a,CoEA2018b,CoEA2018d,CoEA2018c,AhEA2018,CoAh2019} and GW events from LIGO.
In addition to finding the ``afterglow'' associated with a highly relativistic jet powered by a SGRB \citep{WiRe1997,MeRe1998,AsCo2018}, our goal has been to identify a KN, the ultraviolet/optical/near-IR emission generated by the radioactive decay of r-process elements \citep{LaSc1974,LiPa1998,MeMa2010,RoKa2011,Ro2015,KaMe2017}.
The ZTF and Palomar Gattini-IR surveys are our discovery engines, and the Global Relay of Observatories Watching Transients Happen (GROWTH) network\footnote{http://growth.caltech.edu/} is our follow-up network. GROWTH uses a variety of facilities worldwide across various wavelengths 
to perform rapid follow-up and classification of objects.

There are many survey systems participating in the searches for GW counterparts. Amongst many others, the Dark Energy Camera (DECam; \citealt{FlDi2015}), the Gravitational-wave Optical Transient Observer (GOTO; \citealt{Ob2018}), the Panoramic Survey Telescope and Rapid Response System (Pan-STARRS; \citealt{KaBu2010,ChMa2016}), the All-Sky Automated Survey for Supernovae (ASASSN; \citealt{ShPr2014}) and Asteroid Terrestrial-impact Last Alert System (ATLAS; \citealt{ToDe2018}) all have performed observations of events during the third observing run. ZTF provides a competitive addition to these systems, given its depth ($m_\textrm{AB} \sim 20.6$ in 30\,s), wide field of view (FOV $\approx$ 47 deg$^2$ per exposure), and average cadence of $\sim 3$ days over the entire accessible sky. In particular, the cadence is important for establishing candidate history when performing target of opportunity (ToO) observations. The SGRB program, that has covered localization regions spanning thousands of square degrees \citep{CoAh2019}, demonstrated that ZTF is capable of detecting GW170817-like sources out to the Advanced LIGO/Virgo detection horizon at about ($\sim$200\,Mpc; \citealt{AbEA2018}).
In addition, Palomar Gattini-IR (\citealt{Moore2019}, De et al. in prep.) is covering the entire visible northern sky every 2 nights to a $J$-band depth of $\approx 15.5 - 16$ AB mag.
With its 25 deg$^2$ FOV and near-infrared sensitivity, Palomar Gattini-IR provides a complementary system for objects that are expected to be as red as KNe \citep{Me2017}, albeit at lower sensitivity (a source as bright as GW170817 would be detected at $\sim$20\,Mpc). 

The first BNS detection of O3, LIGO/Virgo S190425z, was a single detector event discovered by the Advanced LIGO-Livingston detector, with Virgo also observing at the time \citep{SiEA2019a}.
Occurring at 2019-04-25 08:18:05 UTC, the estimated false alarm rate was 1 in 70,000 years, with a high likelihood of being a binary neutron star.
The first reported BAYESTAR skymap provided an extremely coarse localization, resulting from the low signal-to-noise ratio in Advanced Virgo; it spanned $\sim$\,10,000 deg$^2$, which is nearly a ``pi of the sky.''
The updated LALInference skymap \citep{SiEA2019b}, released at 2019-04-26 15:32:37 UTC, reduced the localization region requiring coverage by $\approx$\,25\% to $\sim$\,7500 deg$^2$.
The all-sky averaged distance to the source is $156 \pm 41$\,Mpc.

In this paper, we describe an $\sim$\,8000 square degree search for the KN counterpart to a single-detector GW event.
Our campaign emphasizes the key role played by both large FOV telescopes like ZTF and Palomar Gattini-IR, as well as the associated follow-up systems.
We demonstrate that our strategy for tiling the sky, vetting candidates, and pursuing follow-up is robust, and capable of promptly reducing 338,646 transient alerts from ZTF to a handful of interesting candidates for follow-up.
Our paper is structured as follows. We describe our observing plan in Section~\ref{sec:observing}.
The identified candidates, including their follow-up, are detailed in Section~\ref{sec:candidates}.
We summarize our conclusions and future outlook in Section~\ref{sec:conclusions}.

\section{Observing Plan}
\label{sec:observing}

\begin{figure}[t]
 \includegraphics[width=3.5in]{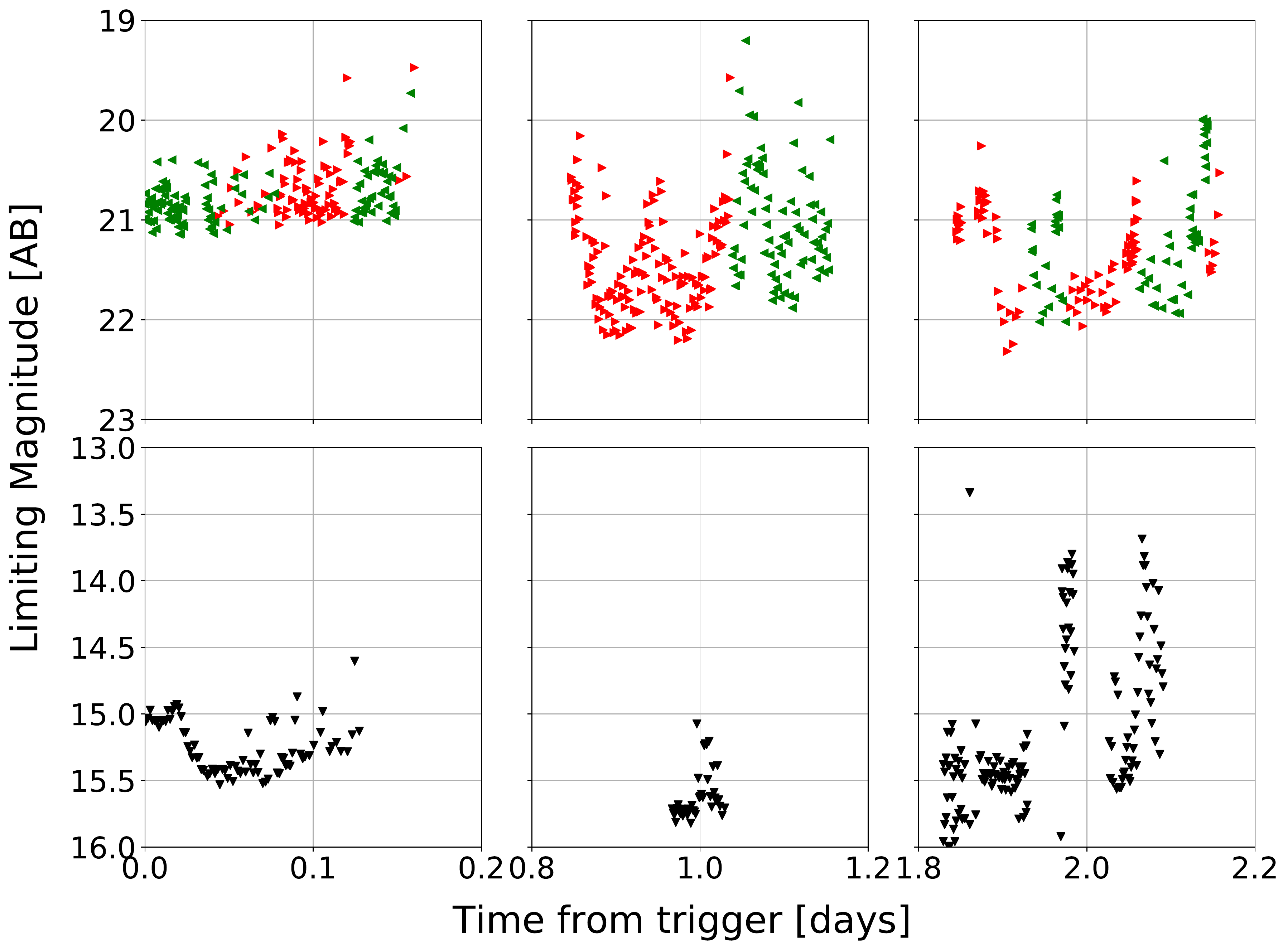}
  \caption{The limiting magnitude as a function of time for S190425z. On the top row is ZTF, while the bottom row is Palomar Gattini-IR, with the left, middle, and right panels corresponding to observations on the first, second, and third nights. The red and green triangles correspond to the $r$- and $g$-band limits from ZTF, while the black triangles correspond to the $J$-band limits from Palomar Gattini-IR.}
 \label{fig:limmag}
\end{figure} 

Because S190425z came during Palomar night-time (2019-04-25 08:18:05 UTC), it occurred concurrently with ongoing survey observations by both ZTF and Palomar Gattini-IR. Within the 90\% localization, approximately 44\% of the original BAYESTAR map was observable from Palomar over the whole night, corresponding to $\approx$ 5000 deg$^2$. The GW event was automatically ingested into the GROWTH ToO Marshal, a database we specifically designed to perform target-of-opportunity follow-up of events localized to large sky-error regions, including GW, neutrino, and gamma-ray burst events \citep{CoAh2019}. Amongst several other features, the ToO marshal allows us to directly trigger the telescope queue for certain facilities to which GROWTH has access, namely ZTF, Palomar Gattini-IR, DECam, Kitt Peak EMCCD Demonstrator (KPED) on the Kitt Peak 84 inch telescope \citep{Coughlin2018}, the Lulin One-meter Telescope (LOT) in Taiwan and the GROWTH-India telescope\footnote{https://sites.google.com/view/growthindia/} (Bhalerao et al., in prep.). We provide a brief description of each instrument in Table~\ref{table:telescopes}.

\begin{table*}
\centering
\caption{Telescope specifications, including name, field of view, pixel scale, telescope aperture, and available filters.}
\label{table:telescopes}
\begin{tabular}{lllllll}
\hline\hline
Name & FOV & Pixel Scale & Aperture  & Filters \\ \hline
ZTF    & 47 deg$^2$ & 1.0$^{\prime\prime}$ & 48\,in & g,r,i \\ 
Palomar Gattini-IR & 25 deg$^2$ & 8.7$^{\prime\prime}$ & 30\,cm & J \\
GROWTH-India &  0.5 deg$^2$ & 0.67$^{\prime\prime}$ & 70\,cm & u,g,r,i,z \\ 
LOT   & 13.2$^\prime$ $\times$ 13.2   $^\prime$ &   0.39$^{\prime\prime}$   &  1\,m  & g,r,i \\
KPED   & 4.4$^\prime$ $\times$ 4.4$^\prime$ & 0.26$^{\prime\prime}$ & 2.1\,m & g,r,U,B,V,I \\
\end{tabular}
\end{table*}

Triggering ToO observations for survey instruments like ZTF and Palomar Gattini-IR halts their ongoing survey observations and redirects them to observe only certain fields as directed by an observation plan.  The observation plan generated by the ToO marshal relies on \texttt{gwemopt} \citep{CoTo2018,CoAn2019}, a code that optimizes the telescope scheduling process for gravitational wave follow-up.  \texttt{gwemopt} handles both synoptic and galaxy-targeted search strategies; we employed the former to conduct observations with some of our facilities, Palomar Gattini-IR, GROWTH-India and ZTF, and the latter for scheduling observations with KPED. The coverage for both ZTF and Palomar Gattini-IR is shown in Figure~\ref{fig:skymap}, and the limiting magnitudes as a function of time in Figure~\ref{fig:limmag}.



\subsection{ZTF}

Serendipitously, after the BNS merger time and before the GW alert was distributed, ZTF had already observed 1920 deg$^2$ of the sky in the $r$-band, corresponding to $\sim19\%$ of the initial BAYESTAR map and $\sim12\%$ of the LALInference map. This overlap between ongoing survey observations and the LIGO-Livingston-only localization is unsurprising as both of the Advanced LIGO interferometers have maximum sensitivity in the sky overhead in North America \citep{FiCh1993, KaNi2014}.  

ZTF triggered ToO observations lasting three hours starting at 2019-04-25
09:19:07.161 UT, one hour after the trigger time. On night~$1$, our observing strategy involved a sequence of $g$-$r$-$g$ band exposure blocks; each exposure was 30\,s, with a typical depth of 20.4 mag, which is the normal duration of exposures during ZTF survey operation. The $g$-$r$-$g$ sequence is the baseline observing strategy for GW follow-up with ZTF as it is specifically designed to capture the inter- and intra-night color evolution of GW170817-like KNe and to distinguish them from supernovae \citep{2017Sci...358.1574S,KiFo2017}.  Due to the size of the localization, we obtained a $g$-$r$ sequence, requiring references for each scheduled field.  In addition, we required a 30 minute gap between observations in $g$ and $r$ to avoid asteroids. Accounting for the loss in probability due to chip gaps and the processing success, ZTF covered 3250 deg$^2$, corresponding to about 36\% of the initial BAYESTAR and 19\% of the LALInference maps on night~$1$.

Motivated by the increase in available observation time ($\sim$\,5 more hours than the first night), we modified our strategy on night~$2$ by taking longer integrations of 90\,s each, corresponding to an average depth of 21.0 mag.  We obtained one epoch in each of $g$- and $r$-band, corresponding to about 46\% probability in the initial BAYESTAR or 21\% of the LALInference maps.

After our observations on both nights were complete, a new LALInference skymap was released at 2019-04-26 14:51:42 UT \citep{gcn24228}.  The LALInference runs reduced the skymap to $\sim$7500 deg$^2$ and shifted more of the probability to two lobes near the sun and in the Southern hemisphere (see Figure~\ref{fig:skymap}).  
In summary, ZTF covered about 8000 deg$^2$ within the 99\% integrated probability region within its two nights of observations.  This corresponds to 46\% of the probability in the original BAYESTAR skymap and 21\% of the probability in the LALInference skymap.  Our observations with ZTF over the two nights covered a 5\,$\sigma$ median depth of $m_\textrm{AB} =$ 21.0 in $r$-band and $m_\textrm{AB} =$ 20.9 in $g$-band.

\subsection{Palomar Gattini-IR}

Palomar Gattini-IR initiated target of opportunity observations of the localization region at 2019-04-25 09:12:09 UT, 11 minutes after the initial notice time.
The synoptic tiling strategy was determined in the same way as for ZTF \citep{CoTo2018}. 
Palomar Gattini-IR imaged a total of 2401 deg$^2$ of the localization region spread over 227 field tiles, covering 32\% of the probability region of the BAYESTAR skymap and 19\% for the LALInference localization.
Each field visit consisted of a sequence of 8 dithered exposures of 8.1\,s each, amounting to a total exposure time of 64.8\,s per field. This resulted in a median stacked depth of $m_\textrm{AB} = 15.5$ in J-band. 
The real-time data reduction pipeline (De et al. in prep) reduced the data and identified transient candidates through the 
application of difference imaging using reference images of the fields.

\subsection{Galaxy Targeted Follow-up}

In addition to the synoptic surveys for counterparts, a subset of the available systems performed galaxy-targeted follow-up. 
This strategy was used by a number of teams to observe GW170817 \citep{ArHo2017,2017Sci...358.1556C,VaSa2017}.
The galaxy-targeted follow-up program relies on the Census of the Local Universe (CLU) catalog \citep{CoKa2017}; it is complete to 85\% in star-formation and 70\% in stellar mass at 200\,Mpc.
The sky area coverage of galaxies is $\approx 1$\,\% within these local volumes \citep{CoKa2017}. 
This makes targeted galaxy pointing tractable for small FOV telescopes (see \cite{ArMc2017} or \cite{GoBu2018} for example). 
Of the galaxies within the volume, our work prioritizes them for follow-up as follows.

The GROWTH ToO Marshal uses an algorithm modified from LCO's galaxy-targeted follow-up of GW events \citep{ArMc2017}, which uses a combination of a galaxy's location in the GW localization region (including the distance), ${S}_{\mathrm{loc}}$, the galaxy's absolute B-band luminosity, ${S}_{\mathrm{lum}}$, and the likelihood of detecting a counterpart at the galaxy's distance ${S}_{\det}$.
We define ${S}_{\det}$ as a prioritization of a transient's potential brightness, taking a fiducial limiting magnitude, ${m}_{\mathrm{lim}}$, for the exposures of $m_\textrm{AB} = 22$, and convert it to a limiting apparent luminosity ${L}_{\mathrm{lim}}$. 
We also compute the luminosity for a potential transient with an absolute magnitude between $-$12 and $-$17, using wide bounds to be robust against differences in intrinsic brightness.
Then, ${S}_{\det}$ becomes ${S}_{\det} = \frac{L_\mathrm{KN max} - L_\mathrm{KN min}}{L_\mathrm{KN max}-L_\mathrm{lim}}$, that we limit to be between 0.01 and 1.
Our final metric is therefore $S = {S}_{\mathrm{loc}} \times {S}_{\mathrm{lum}} \times {S}_{\det}$.

Beginning 4\,hrs after the event, LOT observed 85 galaxies in the initial 90\% localization \citep{gcn24193,gcn24274}. LOT used 180\,s exposures in $R$-band with seeing varying between 1.5-2.5 arcsec. Using comparisons to Pan-STARRS images, these exposures yielded a typical 5\,$\sigma$ limiting magnitude of $m_\textrm{AB} = 20$.
Similarly, KPED started the galaxy targeted follow-up 1.9 hours after the merger and continued until the first ZTF candidates came online. KPED imaged 10 galaxies in the $r$-band filter for 300 seconds, finding no visible transients up to $r=20.8$ \citep{gcn24198}. 300\,s is the fiducial time chosen for KPED to potentially reach limiting magnitudes of $m_\textrm{AB} = 22$, useful for both the transient discovery and follow-up \citep{Coughlin2018}. 

\section{Candidates}
\label{sec:candidates}

We now briefly describe the candidate filtering criteria for the ToO program for ZTF and Palomar Gattini-IR (see \citealt{CoAh2019} for further details).  For GROWTH-India, LOT, and KPED, we did not identify any viable counterparts without previous history of variability in the analysis. 

\subsection{Candidates from ZTF} \label{sec:ztf_cand_description}

A ZTF transient alert is defined as a 5$\sigma$ change in brightness in the image relative to the reference epoch. For ZTF, all transient alerts flagged for follow-up required at least two detections separated by 15 minutes in order to remove asteroids and other transient objects. We used the Pan-STARRS1 point source catalog (PS1 PSC; \citealt{TaMi2018}) to remove candidates located less than 2 arcsec from likely point sources (i.e., stars). Full details on the PS1 PSC can be found in \citet{TaMi2018}; briefly, the authors build a machine learning model that determines the relative likelihood that a PS1 source is a point source or extended based on PS1 colors and shape measurements. The model is trained using sources observed with the \textit{Hubble Space Telescope}, achieving an overall accuracy of $\sim$94\%, and classifying $\sim$1.5$\times 10^{9}$ total sources. 

We also used a real-bogus (RB) classifier to remove common image subtraction artifacts \citep{Mahabal2018}. This method consists of a random forest classifier trained with real objects and artifacts from ZTF images, separating objects with an accuracy of $\sim$89\%. In order to capture the majority of real events, the threshold was set to $RB>0.25$. In addition, the transients must have brightened relative to the reference image, leading to a positive residual after the image subtraction. Furthermore, the program excluded all objects within 20 arcsec of $m_\textrm{AB} < 15$ stars to avoid artifacts from blooming, thus excluding $\sim2-5$\% of the imaged region, which depends significantly on stellar density.\footnote{Estimates of the amount of excluded area rely on the assumption that the sky fraction excluded around $m_\textrm{AB} < 15$ stars, within a few circular regions of 1 deg$^2$ in the skymap that we checked, is representative of the overall sky fraction excluded from the entire imaged region.}
The final step involved constraining the search to events that have no historical detections prior to three days before the trigger. 

This filtering scheme reduced the number of ZTF alerts from 50802\, to 28\, for the first night and from 287844\, to 234\, relevant candidates for the second night. A more detailed breakdown on the number of alerts that successfully met the criteria at each  filtering step can be found in Table \ref{table:filtering_ztf}.
\begin{table}[ht]
\centering
\caption{Filtering results for both ZTF nights. The quantities represent the number of alerts that passed a particular step in the filter. Each step is run over the remaining alerts from the previous stage. The criteria are described in Section \ref{sec:ztf_cand_description} and the total number of relevant candidates is highlighted. In particular, ``Real'' indicates a real-bogus score greater than 0.25, and ``not moving'' indicates that are there more than 2 detections separated by at least 30 minutes.}
\label{table:filtering_ztf}
\begin{tabular}{lp{19mm}p{19mm}}
\hline \hline
Filtering criteria & \# of Alerts on April-25 & \# of Alerts on   April-26   \\ \hline 
ToO alerts & 50,802 & 287,844 \\
Positive subtraction & 33,139 & 182,095\\ 
Real & 19,990 &118,446\\ 
Not stellar & 10,546 &61,583\\ 
Far from a bright source & 10,045 &58,881\\ 
Not moving & 990 &5,815\\ 
No previous history & \textbf{28} & \textbf{234}\\ \hline 

\end{tabular}
\end{table}

The candidates that passed these criteria were filtered and displayed by the GROWTH Marshal \citep{Kasliwal2018}, a database used to display historical lightcurves (including upper limits) for each object that also performs cross-matches with external catalogs. We subjected each of the remaining candidates to a thorough human vetting process to determine whether the transient could be a viable counterpart to S190425z. Through this vetting process, we removed candidates whose coordinates were outside the 90\% contour in the GW localization, and candidates that had archival detections in the Pan-STARRS1 Data Release 2 \citep{Flewelling2018}.  We flagged Active Galactic Nuclei (AGN) based on the WISE colors \citep{WrEi2010} for each transient and its offset from the nucleus of the galaxy.  Furthermore, we prioritized candidates whose photometric/spectroscopic redshift was consistent with the GW distance estimate, and whose extinction-corrected lightcurve exhibited rapid color evolution initially.  For the most promising candidates in our vetted list, we performed forced photometry at the position of the source to ensure there were no historical detections with ZTF.

Our first night of observations yielded only two such candidates that passed both the automatic filtering and human vetting processes. These two candidates were ZTF19aarykkb and ZTF19aarzaod. The second night of observations allowed us to identify additional candidates detected on the first night that were consistent with the new skymap, thereby increasing our candidate list from two to 13 from the first night to the second. We describe the most promising of these 15 candidates in more detail in Sec. \ref{sec:ztf_candidates}.     

To double-check that we did not miss any candidates, we used \texttt{Kowalski}\footnote{\url{https://github.com/dmitryduev/kowalski}}, an open-source system used internally at Caltech (primarily) to archive and access ZTF's alerts and light curves \citep{DuMa2019}. Specifically, we used \texttt{Kowalski}'s web-based GUI called the ZTF Alert Lab (ZAL), with which users can efficiently query, search and preview alerts. Our results were consistent with the results above. 
To triple-check that we did not miss any candidates, we also carried out an additional automatic search of the AMPEL alert archive \citep{Nordin:2019kxt} for transients that might have escaped. No additional candidates from either night were found. 

\subsection{Candidates from Palomar Gattini-IR}
\label{sec:candidates_gattini}
For Palomar Gattini-IR, we adopted the following selection criteria for human vetting of sources identified in the difference imaging:
\begin{enumerate}
    \item We selected candidates that were at least 1\,arcminute away from bright stars with $m_\textrm{J} < 10$, excluding $\sim0.7-2$\% of the imaged region, in order to remove contamination from subtraction artifacts.\footnote{Estimates of the amount of excluded area rely on the assumption that the sky fraction excluded around $m_\textrm{AB} < 10$ stars, within a few circular regions of 1 deg$^2$ in the skymap that we checked, is representative of the overall sky fraction excluded from the entire imaged region.}
    \item The first detection of the candidate must have been after the gravitational-wave trigger time.
    \item An object must have at least two detections with a signal-to-noise ratio greater than 5 or a signal-to-noise ratio greater than 7 in one detection. Amongst sources with single detections, we also rejected known asteroids.
\end{enumerate}

No viable counterparts were identified in this search.

\subsection{Follow-up of ZTF candidates}
\label{sec:ztf_candidates}

The 15 sources that were identified from ZTF observations are shown in Table~\ref{table:followup} and on Figure~\ref{fig:skymap}. Using a variety of resources including the SED Machine (SEDM) \citep{nblago18,RiNe2019} on the Palomar 60 inch (P60) telescope, the Double Beam Spectrograph (DBSP; \citealt{OkGu1982}) on the Palomar 200 inch (P200) telescope, the Robert Stobie Spectrograph (RSS; \citealt{Smith2006}) on the Southern African Large Telescope (SALT), the Liverpool telescope (LT; \citealt{StSm2004}), the GROWTH-India telescope, 
the KPED, the Himalayan Chandra Telescope (HCT), the Discovery Channel Telescope (DCT) and LOT, we followed up each of these candidates with further photometry and/or spectroscopy.

A total of 5 objects were classified using spectroscopy \citep{gcn24321,gcn24204,gcn24205} and we tracked the color evolution of 15 objects using photometry for about 7 days on average. A KN is expected to show a rapid evolution in magnitude \citep{Me2017}; GW170817 faded $\Delta r \sim 1$\,mag per day over the first 3 days and by $\Delta r \sim 4.2$\,mags total around day 10. Thus, we can use photometric lightcurves to determine whether a transient is consistent with the expected evolution for a KN. Some photometrically monitored transients showed evolution that was too slow ($\Delta r \sim 0.1$\,mag per day) to be consistent with GW170817 or kilonova model predictions.
Many other candidates highlighted in \citealt{gcn24191} were observed with GROWTH facilities, however, they were later excluded by the updated LALInference skymap. In addition to these sources, we reported objects in \citealt{gcn24191} with ZTF detections before the event time to the community in order to limit the number of false positives identified by other surveys that may not have recently imaged those areas of the sky.


\begin{figure*}[!htb]
  \begin{minipage}[b]{0.5\linewidth}
    \centering
    \includegraphics[width=\linewidth]{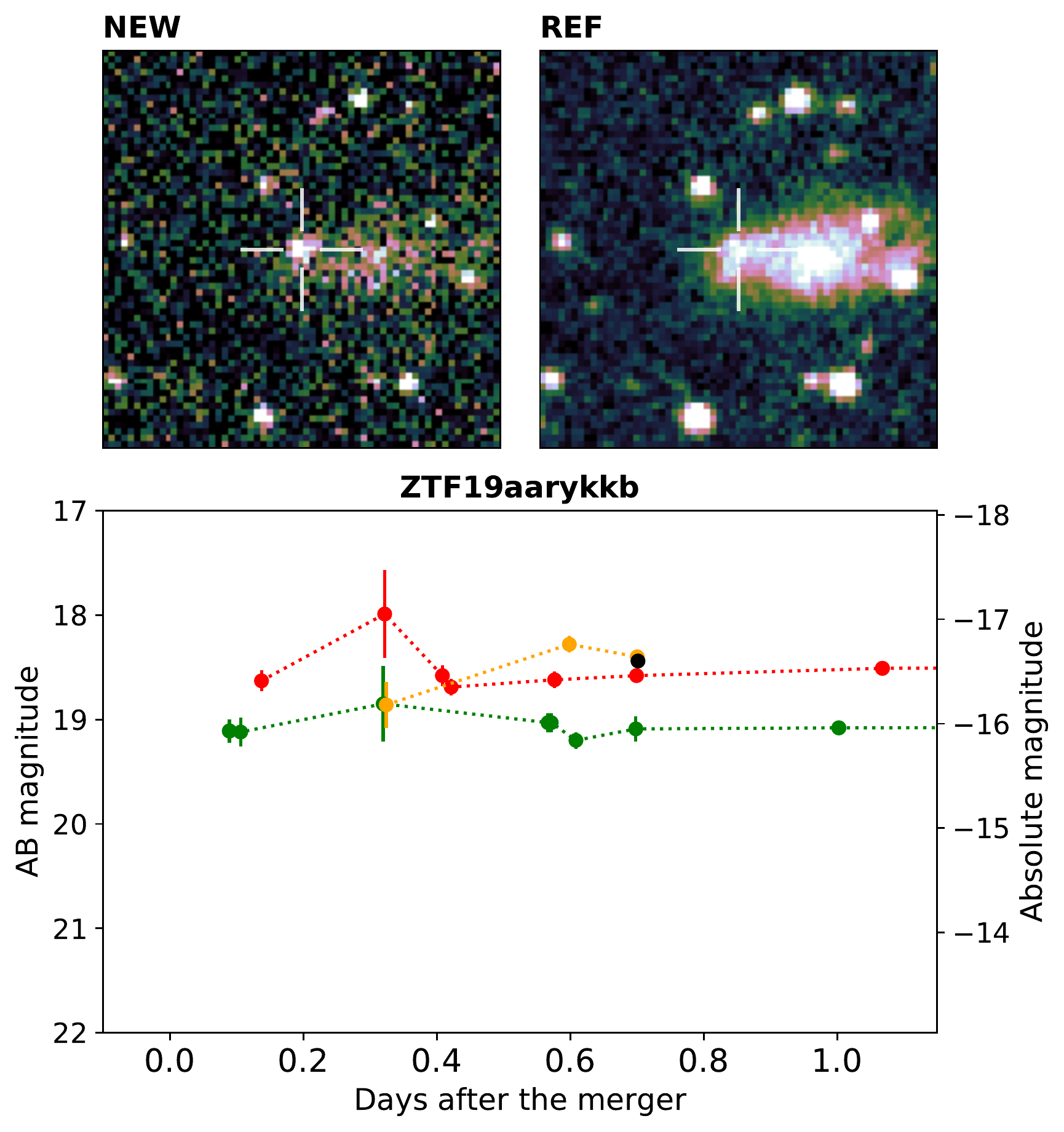}

  \end{minipage}
  \begin{minipage}[b]{0.5\linewidth}
    \centering
    \includegraphics[width=\linewidth]{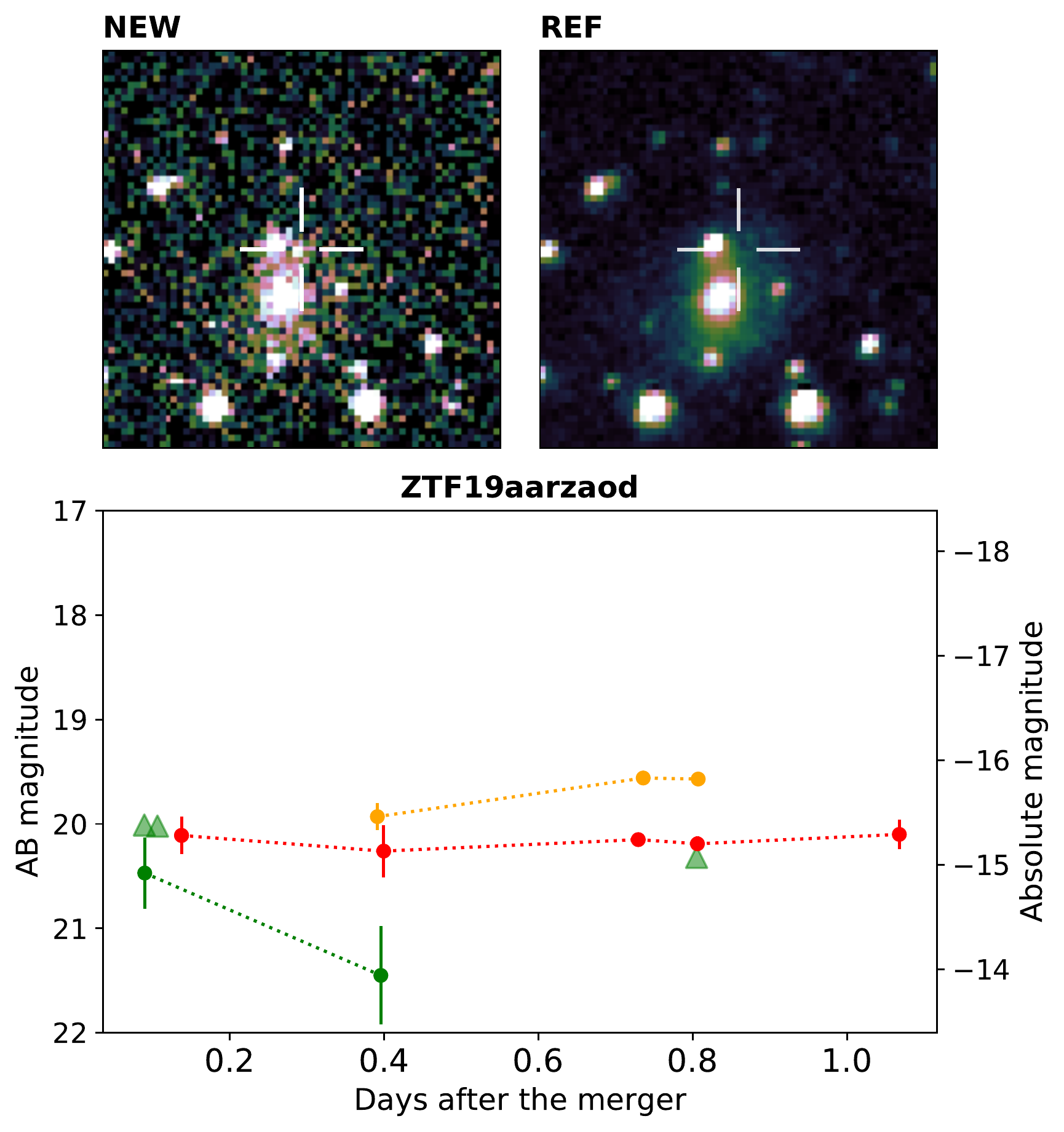}
  \end{minipage} 
  \begin{minipage}[b]{0.5\linewidth}
    \centering
    \includegraphics[width=\linewidth]{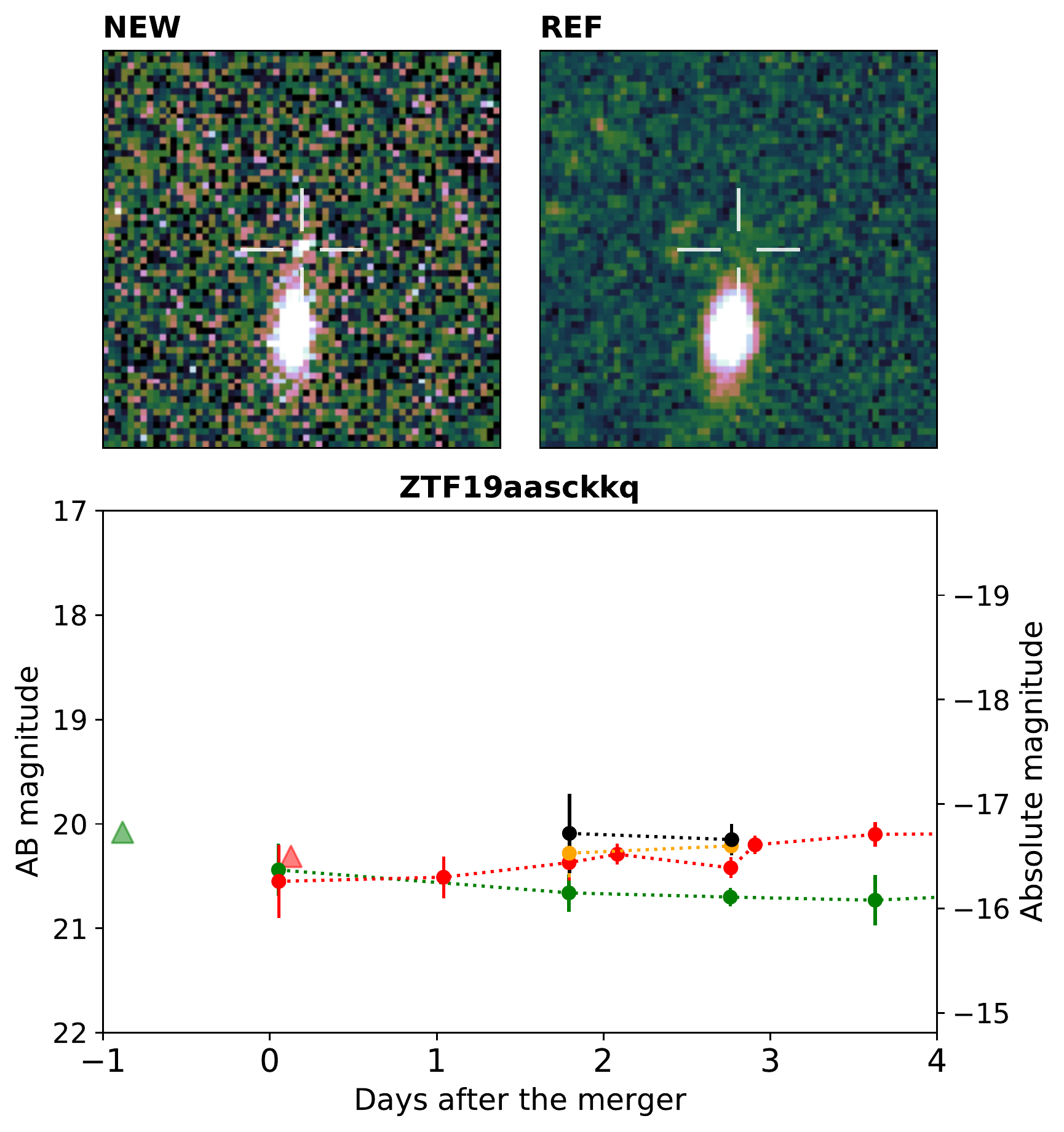}
  \end{minipage}
  \begin{minipage}[b]{0.5\linewidth}
    \centering
    \includegraphics[width=\linewidth]{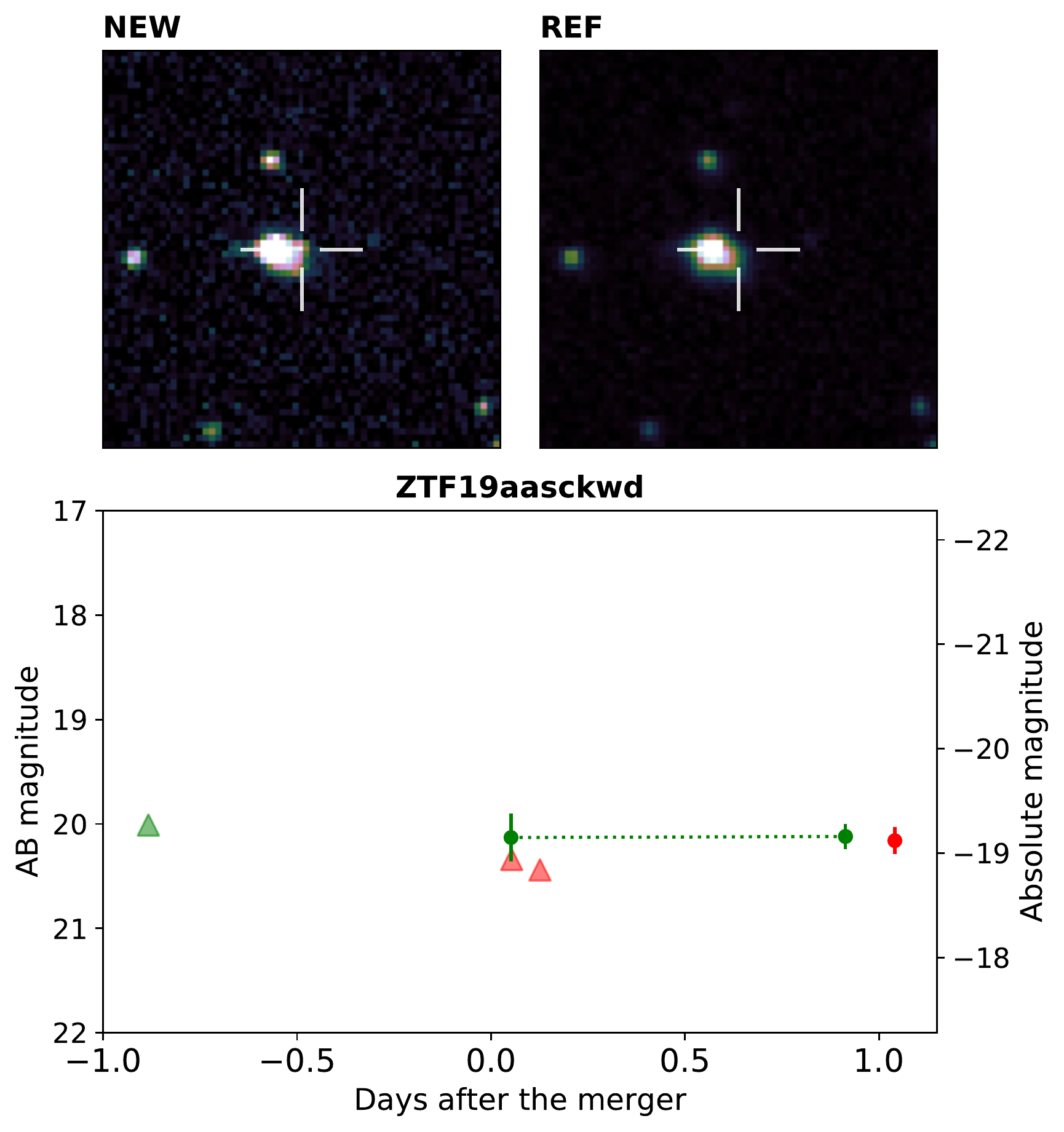}
  \end{minipage} 

\caption{Lightcurves and r-band cutouts for the ZTF candidates discussed in Section \ref{sec:ztf_candidates}. The lightcurves are constructed with data acquired with GROWTH facilities: for ZTF19aarykkb, the data is from ZTF, LOT, GIT and LT, for ZTF19aarzaod, ZTF, LOT and LT, for ZTF19aasckkq, ZTF, KPED and LT and for ZTF19aasckwd, ZTF and KPED. We used colors to represent each band in the lightcurves: green for g-band, red for r-band, yellow for i-band and black for z-band. While triangles in the lightcurve represent upper limits, filled circles are the magnitudes of the object. For each transient, the cutout on the left corresponds to the ZTF discovery image and the right cutout corresponds to the ZTF reference image of the host. A cross marks the location of the transient in the reference image. The cutouts are 0.7 sq. arcmin with north being up and east to the left.}\label{fig:lc_transients}
\end{figure*}

We now provide a broad summary of the most promising candidates ruled out by spectroscopy, as examples of the follow-up performed by the GROWTH facilities when vetting candidates. In particular, we highlight the lightcurves of ZTF19aarykkb, ZTF19aarzaod, ZTF19aasckkq, and ZTF19aasckwd in the top left, top right, lower left and lower right panels respectively
in Figure~\ref{fig:lc_transients} and discuss them briefly below. The associated spectra are shown in the top panel of Figure~\ref{fig:transients}; the spectrum of ZTF19aasckwd is not shown as we only have a spectrum of the galaxy host. We used the value of $H_0=67.4$\,km $\mathrm{s}^{-1}$ $\mathrm{Mpc}^{-1}$ \citep{AgEA2018} to calculate absolute magnitudes.

\subsubsection{ZTF19aarykkb} 
We first detected the transient ZTF19aarykkb 2.13\, hours after the merger and highlighted it in the first ZTF GCN \citep{gcn24191}. ZTF19aarykkb is 12.1\, arcsec offset from the host galaxy, which is at a redshift of $z=0.024$, corresponding to a luminosity distance of 106 Mpc. The absolute magnitude of the discovery is $g=-15.9$, broadly consistent with GW170817 and KNe predictions. 
We ran forced photometry in archival ZTF images of the region, finding no variability at the coordinates before the merger. The last upper limit at this location was 5.8\, days before the LVC alert in $g$-band ($m_\textrm{AB} > 18.74$ in $g$-band). 
Due to its distance and discovery mag, several facilities followed-up this source \citep{gcn24204, gcn24206, gcn24219, gcn24220, gcn24226, gcn24321, gcn24260} The LOT group in Taiwan imaged the object 6 hours after the transient set in Palomar \citep{gcn24193}; later that day, the LT continued the monitoring. This object was imaged 18 times within the first 26 hours after the merger. The first spectrum for this object came from the Himalayan Chandra Telescope (HCT) about 10.67 hours after the trigger \citep{gcn24200}, showing a strong H$\alpha$ line at a redshift of $z=0.024$. This was confirmed 8 hours later by the LT team with the Spectrograph for the Rapid Acquisition of Transients (SPRAT) \citep{piascik2014sprat}, who classified it as a young SN Type II \citep{gcn24204}, based on the characteristic P-Cygni profile in the LT spectrum. An additional spectrum was taken about 10 hours later with the DeVeny spectrograph mounted on the 4.3\,m DCT \citep{gcn24220}, showing similar strong H$\alpha$, furthermore confirming the SN classification (see Figure \ref{fig:transients}).


\subsubsection{ZTF19aarzaod} 

ZTF19aarzaod was first detected by ZTF 2.15\,hrs after the merger \citep{gcn24191} with its last upper limit ($m_\textrm{AB} > 20.01$ in $g$-band) 6 days prior the merger. Forced photometry did not show previous history of variability at the transient location. The redshift of the host galaxy is $z=0.028$, putting the transient at a distance of 128.7\, Mpc. The transient is offset by 8.2\,arcsec from the host galaxy and its absolute magnitude at discovery was $r=-15.3$, also consistent with a GW170817-like KN. 
ZTF19aarzaod was extensively followed-up with various observatories \citep{gcn24194, gcn24205, gcn24208, gcn24209, gcn24214, gcn24219, gcn24226, gcn24321} and was imaged 13 times during the first day.
Spectroscopic observations of ZTF19aarzaod were taken with RSS mounted on SALT on UT 2019-04-26.0 under a special gravitational-wave follow-up program 2018-2-GWE-002 and reduced with a custom pipeline based on PyRAF routines and the PySALT package \citep{Crawford2010}. 
The spectrum covered a wavelength range of 470-760\,nm with a spectral resolution of R = 400.
The spectrum shows broad H$\alpha$ emission along with some He I features (see Fig.~\ref{fig:transients}) classifying it as a type II supernova at $z = 0.028$ \citep{gcn24205}.

\subsubsection{ZTF19aasckkq} 

The transient ZTF19aasckkq \citep{gcn24311} was first detected by ZTF 1.23\,hrs after the merger. It is offset from the host galaxy by 10.1\,arcsec, and its last upper limit ($m_\textrm{AB} > 20.1$ in $g$-band) was the night before the merger. We ran forced photometry at the location of the transient, finding no activity before the merger. The discovery absolute mag is $r=-16.3$, similar to GW170817 at peak. ZTF19aasckkq was followed-up 18 hours after the last ZTF detection by LT and KPED \citep{gcn24320}. This transient was imaged 16 times for a period of 3.8\,days by a variety of observing groups \citep{gcn24314, gcn24319, gcn24320, gcn24343}. \citealt{gcn24321} first classified ZTF19aasckkq as a Type IIb SN at z$\sim$0.05, consistent with the galaxy redshift \citep{HoCo2019}.  In Figure~\ref{fig:transients}, we highlight the presence of He I, H$\alpha$ and H$\beta$ absorption features in the first spectrum we acquired with P200+DBSP, confirming its classification as a SN IIb at a redshift of $z=0.0528$. The source was still bright at $r=19.8$, 14 days after S190425z.

\subsubsection{ZTF19aasckwd} 
ZTF19aasckwd was detected 1.23\,hrs after the merger about 4.2\,arcsec from its host galaxy \citep{gcn24311}. Its last upper limit ($m_\textrm{AB} > 20.1$ in $g$-band) was the night before the trigger. The forced photometry search did not show activity prior to the merger. This transient was imaged 5 times during the first 24\,hrs and it was classified as a SN Ia by \cite{gcn24321} at a redshift of $z=0.145$ \citep{HoCo2019}. The absolute magnitude at discovery was $r=-19.2$, a few magnitudes brighter than what is expected from a KN.

\subsection{Follow-up of non-ZTF candidates}
Here, we report on the follow-up triggered by the GROWTH team of a number of transients discovered by other facilities to be consistent with the LALInference skymap. We queried the GROWTH follow-up marshal at the positions of the most promising transients announced in order to determine whether 1) the transient had historical detections with ZTF, or 2) our concurrent photometry of the object also supported the KN hypothesis. Additionally, we used LT, GROWTH-India Telescope, and DECam to obtain photometry of the candidates that were not detected with ZTF because they were either fainter than the ZTF average upper limits or inaccessible due to their sky location. Table \ref{table:followup_nonztf} summarizes the most relevant non-GROWTH objects followed-up by the GROWTH collaboration, and we briefly discuss them below.

\subsubsection{\emph{Swift}'s Ultraviolet/Optical Telescope (UVOT) candidate}\label{sec:uvot}
\begin{figure}
    \centering
    \includegraphics[width=3.6in]{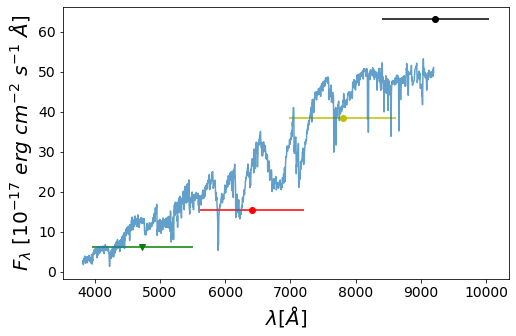}
    \caption{The DECam (g, r, i and z-band) fluxes of the UVOT candidate discussed on Section \ref{sec:uvot} are over-plotted on the spectra of an SDSS M2-dwarf.}
    \label{fig:decam_uvot}
\end{figure}

We followed up photometrically the \emph{Swift}/UVOT candidate \citep{gcn24296}, discovered at RA=17:02:19.2, Dec=$-$12:29:08.2 in $u$-band with $m_\textrm{Vega}=17.7 \pm 0.2$.
The transient was within a few hundred arcseconds of two galaxies within the localization volume. After its initial detection with Swift, several other facilities \citep{gcn24296, gcn24301, gcn24302, gcn24304, gcn24306, gcn24307, gcn24313, gcn24318, gcn24324, gcn24325, gcn24334, gcn24335, gcn24328, gcn24459}, including ZTF and Palomar Gattini-IR, reported non-detections or pre-discovery upper limits that indicated the transient might be rapidly fading in the ultraviolet. \citealt{gcn24312} reported an object offset by $<$ 1 arcsec from the position of the reported UVOT candidate after visually inspecting archival DECam optical images. Using the GROWTH-DECam program, \citealt{gcn24337} detected a source consistent with the coordinates reported by \citealt{gcn24312}, but no transient at the coordinates reported by Swift \citep{gcn24301} (see Table \ref{table:followup_nonztf}). 
The slight trailing observed in images of the original UVOT source (which introduced uncertainty in the astrometry) strongly hinted at the physical association between the transient and the offset source. The colors of the associated source ($r-z=1.53$ and $ g-r>0.97$) are consistent with those of a M2-dwarf \citep{West11}. For this reason, a likely explanation for the observed ultraviolet transient is that it was a galactic M2-dwarf flare \citep{gcn24326,gcn24337}, unassociated with the GW event. The photometry of the UVOT candidate is shown with a SDSS spectra of a M2-dwarf in Figure \ref{fig:decam_uvot}.

\subsubsection{AT2019ebq/PS19qp}
\begin{figure*}[t]
 \includegraphics[width=7.0in]{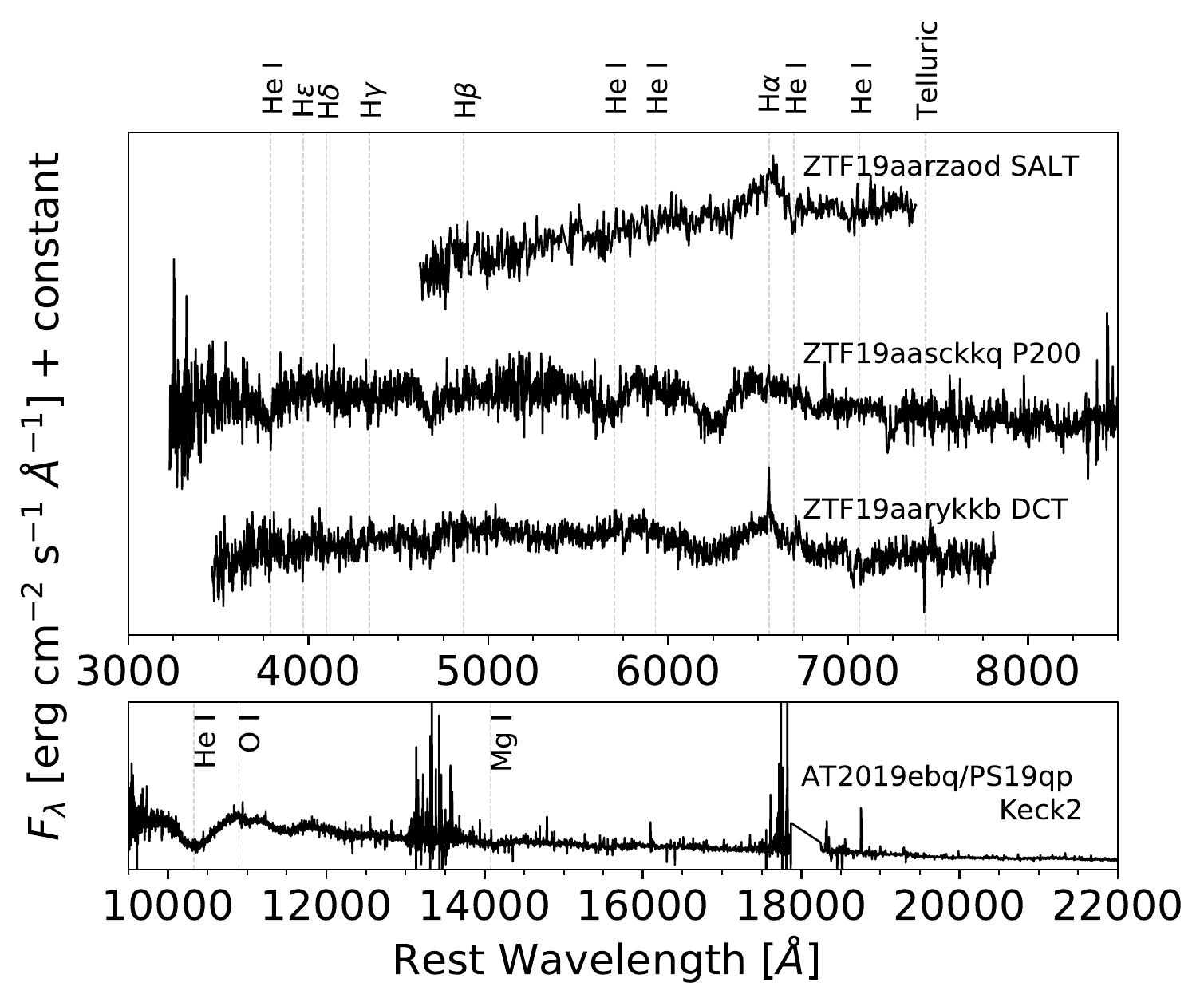}
  \caption{Spectra of all the candidates for which spectroscopic data were taken. The transient name and instrument used to obtain the spectrum are noted on the right hand side of the plot. We show the spectrum for AT2019ebq/PS19qp in its own panel given the different wavelengths covered from the other transients. The dotted gray lines show the characteristic features in each spectrum that helped with its classification. These four transients were all classified as core-collapse SNe.  The classification and phase for each transient is as follows: ZTF19aasckkq - SN IIb, 7 days; ZTF19aarykkb - SN II, 1 day \citep{gcn24220}; ZTF19aarzaod - SN II, 0 days \citep{gcn24205}; AT2019ebq/PS19qp - SN Ib/c, 1 day \citep{gcn24233}.
  }
 \label{fig:transients}
\end{figure*}

We also obtained spectroscopy of AT2019ebq/PS19qp \citep{gcn24210} with the Near-Infrared Echellete Spectrometer (NIRES) on Keck II. This candidate was initially claimed to be exceptional in that its optical spectrum taken with the Gran Telescopio Canarias (GTC) contained broad absorption features ``unlike normal supernovae;'' therefore \cite{gcn24221} highlighted it as a promising KN candidate. Our NIR spectrum taken $\sim 1.5$ days after the trigger, however, exhibited broad P Cygni SN-like features of He I that indicated that the transient was a Type Ib/c SN \citep{gcn24233}, ruling out its association with S190425z (see bottom panel of Fig. \ref{fig:transients}). Several other facilities that also followed up this source helped verify its classification \citep{gcn24229, gcn24230, gcn24233, gcn24295, gcn24241, gcn24252, gcn24358}.

7 additional PS1 candidates (out of the 20 transients reported by \cite{gcn24210}) were ruled out based on previous ZTF detections (\citealt{gcn24349}; see Table~\ref{table:followup_nonztf}).

\subsubsection{Marginal ATLAS candidates}
Additionally, we acquired a short sequence (40 seconds each in $gri$ filters) of imaging at the locations of all five of the marginal ATLAS transients reported by \cite{gcn24197} using IO:O on the 2\,m Liverpool Telescope \citep{gcn24202}. No significant source was detected at the location of any of them (to typical depths of 22\,mag; see Table  \ref{table:followup_nonztf}).  Combined with the fact that none of these transients had a detectable host galaxy, this suggests these transients were likely to be spurious or perhaps short-timescale flares from faint stars.


\begin{table*}
\centering
\caption{Follow-up table for the Palomar Gattini-IR candidate described in Section~\ref{sec:candidates_gattini} and the 15 most interesting ZTF candidates from \cite{gcn24191} and \cite{gcn24311}. The sources with a star (*) have photometric evolution (in units of mag/day) inconsistent with the evolution of a KN (Section~\ref{sec:ztf_candidates}). Spectra obtained with SOAR \citep{gcn24321} were critical in classifying ZTF19aasckwd and ZTF19aasckkq while spectra from SALT \citep{gcn24205} allowed the classification of ZTF19aarzaod. GROWTH teams acquired spectra of ZTF19aarykkb with HCT, LT, and DCT \citep{gcn24200, gcn24204, gcn24220} and also provided useful photometric data towards the classification of these transients \citep{gcn24314, gcn24320, gcn24201, gcn24198, gcn24193}. We monitored the transients on average for 7 days. The redshift, spectroscopic (s) or photometric, (p) of the host galaxy is also listed.}
\label{table:followup}
\begin{tabular}{lllllll}
\hline\hline
Candidate &Coordinates (RA, Dec) & Discov. Mag. &Classification  &Spec. facilities& Phot. evol.&Redshift/Host \\ \hline
ZTF19aarykkb  & 17:13:21.95 $-$09:57:52.1 &  r = 18.63 & SNII z=0.024& HCT, LT, DCT &...& 0.024 (s)\\
ZTF19aarzaod  & 17:31:09.96 $-$08:27:02.6 & r = 20.11 & SNIIn z=0.028 & SALT & ...& 0.028 (s)\\
ZTF19aasckwd  & 16:52:39.45 +10:36:08.3 &  r = 20.15& SN Ia z=0.145 & SOAR  &...& 0.15 (s)\\
ZTF19aasckkq  & 16:33:39.14 +13:54:36.7 &  g = 20.86& SN IIb z=0.052& P200, SOAR  &...& 0.053 (s)\\ 

ZTF19aasbphu  & 16:22:19.95 +21:24:29.5 &  r = 19.71 &Nuclear*&...& 0.11 & 0.0971 (p) \\ 
ZTF19aaryxjf  & 16:58:22.87 $-$03:59:05.1 &  g = 19.95 &SN*&... & -0.014 & 0.07791 (s, GLADE)  \\ 
ZTF19aarxxwb  & 19:14:46.40 $-$03:00:27.0 &  g = 18.89 &SN*&...      & 0.12 & hostless   \\ 
ZTF19aasdajo  & 16:57:25.21 +11:59:46.0 &  g = 20.7 &SN*&...  & 0.045 & 0.292 (p)  \\ 
ZTF19aasbamy  & 15:25:03.76 +24:55:39.3 &  g = 20.66&SN*&...& 0.01 & 0.201 (p) \\ 
ZTF19aarycuy  & 16:16:19.97 +21:44:27.4 &  r = 20.07 &SN*&...       & 0.02 & 0.127 (p) \\

ZTF19aasbaui  & 15:40:59.91 +24:04:53.8 &  g = 20.49 &SN*&...       & 0.01 & 0.04 (s, CLU)  \\ 
ZTF19aasejil  & 17:27:46.99 +01:39:13.4 &  g = 20.53 &SN*&...   & 0.01 & 0.199 (p)\\
ZTF19aascxux  & 17:13:10.39 +17:17:37.9 &  g = 20.56&SN*&...    & 0.06 & 0.165 (p)\\
ZTF19aashlts  & 16:52:45.01 $-$19:05:38.9 &  r = 19.95&SN*&...      & 0.03 & hostless \\ 
ZTF19aasfogv  & 17:27:22.32 $-$11:20:01.9 &  g = 20.53&SN*&...   & 0.01 & hostless\\ \hline

\end{tabular}
\end{table*}

\begin{table*}
\centering
\caption{GROWTH follow-up table for candidates reported by other surveys. GROWTH-India, LOT, and DECam-GROWTH follow-up of the Swift/UVOT candidate discovered by \cite{gcn24296} helped confirm its classification as a likely M-dwarf flare \citep{gcn24296, gcn24301, gcn24302, gcn24304, gcn24306, gcn24307, gcn24312, gcn24313, gcn24318, gcn24324, gcn24325, gcn24326, gcn24334, gcn24335, gcn24337, gcn24328, gcn24459}.  Our initial Keck spectrum of another promising candidate, AT2019ebq/PS19qp \citep{gcn24210} showed it was a Type II SN \citep{gcn24233}. Several of the PS1 candidates reported by \cite{gcn24210}, as well as Gaia19bpt \citep{gcn24354} were found to have previous detections with ZTF \citep{gcn24356, gcn24223}. For these sources, we list the number of days before S190425z that they were detected in parentheses. LT provided constraining upper limits of some reported ATLAS candidates \citep{gcn24197, gcn24202}. }
\label{table:followup_nonztf}
\begin{tabular}{lllll}
\hline\hline
Candidate & Coordinates (RA, Dec) & Discovery Mag. & GROWTH follow-up & upper limits    \\ \hline
UVOT  & 17:02:19.21 $-$12:29:08.2&  u=17.74 & GIT, LOT, DECAM & DECam g $>$ 24.0   \\ 
  ...& ...& ...& ...& DECam r $>$ 24.0 \\
  ...& ...& ...& ...& DECam i $>$ 23.7 \\
  ...& ...& ...& ...& DECam z $>$ 23.1 \\\hline
AT2019ebq-PS19qp & 17:01:18.33 $-$07:00:10.4 & i= 20.40 & Keck spectrum SN Ib/c & ... \\ \hline

Gaia19bpt  & 14:09:41.88 +55:29:28.1 & o = 18.49 & ZTF19aarioci (4.12) & ...\\ \hline

AT2019ebu-PS19pp &14:19:49.43  +33:00:21.7& i =  20.77  & ZTF19aasbgll (2.10) & r=20.60 \\ 
AT2019ebw-PS19pq &15:02:17.02  +31:14:51.6& i =  20.92  & ZTF19aasazok (11.95) & g=20.91 \\ 
AT2019ecc-PS19pw &15:26:29.53  +31:39:47.5& i =  20.10  & ZTF19aapwgpg (17.96) & r=20.14\\
AT2019eck-PS19qe &15:44:24.53  +32:41:11.0& i =  20.81  & ZTF19aapfrrw (24.97) & g=20.13\\
AT2019ecl-PS19qg &15:48:11.85  +29:12:07.1& i =  20.51  & ZTF19aasgwnp (25.89) & g=21.02\\
AT2019ebr-PS19qj &16:35:26.48  +22:21:36.4& i =  19.79  & ZTF18aaoxrvr (25.86) & g=20.83\\
AT2019ebo-PS19qn &16:54:54.71  +04:51:31.5& i =  20.02  & ZTF19aarpgau (9.87) & g=20.40\\ \hline

AT2019eao-ATLAS19hyo &13:01:18.63 +52:09:02.1 & o =  19.36 & LT & g $>$ 22.1 \\ 
AT2019ebn-ATLAS19hwh &13:54:47.42 +44:46:27.3 & o =  19.07 & LT & g $>$ 22.1 \\ 
AT2019ebm-ATLAS19hwn &12:59:58.58 +29:14:30.7 & o =  19.42 & LT & g $>$ 22.3 \\ 
AT2019ebl-ATLAS19hyx &14:32:31.53 +55:45:00.1 & o =  19.28 & LT & g $>$ 22.3 \\ 
AT2019dzv-ATLAS19hxm &14:01:45.02 +46:12:56.1 & o =  19.23 & LT & g $>$ 22.2 \\ \hline
\end{tabular}
\end{table*}

\section{Conclusions}
\label{sec:conclusions}
In this paper, we have described the first follow-up of a binary neutron star event with ZTF and Palomar Gattini-IR. 
Covering more than 8000 deg$^2$ with ZTF and 2400 deg$^2$ with Palomar Gattini-IR over two nights, we show how these systems in combination with follow-up facilities are capable of rapidly identifying and characterizing transients on hour to day timescales over sky regions of this size. 
We show how it is possible to reduce 338,646 alerts to 15 previously unidentified candidate counterparts.
We also show how with the follow-up resources available to GROWTH, we can rule out these objects as viable candidates.

Assuming an optical/NIR counterpart with a luminosity similar to that of GW170817, which had an absolute magnitude of about $-16$ in g, r, and J-bands, the apparent magnitude in these bands for the distribution of distances to S190425z is $m_\textrm{AB}\approx19-20.5$.
This varies between 1\,mag brighter than to near the detection limit for ZTF for this analysis, indicating ZTF is well-primed for detecting a GW170817-like source at these distances.
We expect that a closer or brighter than expected source (GW170817 would be detected at $\sim$20\,Mpc) should be detectable with Palomar Gattini-IR.

As a cross-check of the number of sources we are identifying, we compare to the fiducial supernova rate of $\approx\,10^{-4} {\rm Mpc}^{-3} {\rm yr}^{-1}$ \citep{LiLe2011}. The 90\% localization volume of the gravitational-wave skymap is $\sim$\,$2.1 \times 10^{7} {\rm Mpc}^{3}$. As stated above, ZTF covered about 46\% of the skymap, meaning we expect to detect $\sim$\,$2.1 \times 10^{7} {\rm Mpc}^{3} \times 1.04 \times 10^{-4} {\rm Mpc}^{-3} {\rm yr}^{-1} \times 0.46 \approx 2.7 {\rm day}^{-1}$. Since the distribution of Type II SNe at peak luminosity falls between absolute magnitudes of $\approx$ -15 to -20 mags \citep{Richardson2014}, brighter than the expected distribution at peak for KNe, our follow-up observations with ZTF should have detected all of the bright, and most of the dim Type II SNe.  Having taken images for about 12\,hrs during the nights, we would expect to detect $\sim$\,1-2, consistent with the 2 young supernovae highlighted in this paper.

Going forward, prioritizing further automatized classification of objects can lead to more rapid follow-up and dissemination of the most interesting objects. 
For example, the inclusion of machine-learning based photometric classification codes such as RAPID \citep{MuNa2019} will help facilitate candidate selection and prioritization.
We are also actively improving the scheduling optimization, and have since added a feature to schedule using the ``secondary'' ZTF grid, that is designed to fill in the chip gaps.

The follow-up of S190425z highlights two important points. 
The first is that rapid dissemination of updated GW skymaps is useful for tiling prioritization.
This helps mitigate the effects of shifting localization regions, including potentially decreasing sky areas.
The second is that we are capable of performing nearly all-sky searches with ZTF and Palomar Gattini-IR and conducting the necessary follow-up with partner facilities, even in the case of a single-detector GW trigger.
This event serves to extend the frontier in searches for optical transients in large areas. 
The intermediate Palomar Transient Factory found optical counterparts to eight long GRBs localized to $\sim$\,100 deg$^2$ \citep{SiKa2015}, with GRB 130702A \citep{SiCe2013} being the first of its kind, and this event has shown it is possible to cover more than an order of magnitude larger sky area.
One caveat to this conclusion is that in general, single-detector localizations will include regions on the sky not accessible to one ground-based facility alone; this motivates the use of coordinated networks of telescopes with worldwide coverage \citep{NiKa2013,KaNi2014}.
However, we have demonstrated that the network on hand is capable of overcoming the challenges of rapidly and efficiently searching for electromagnetic counterparts in this new era of gravitational-wave astronomy.

\acknowledgments

We would like to thank Peter Nugent for comments on an early version of this paper.

This work was supported by the GROWTH (Global Relay of Observatories Watching Transients Happen) project funded by the National Science Foundation under PIRE Grant No 1545949. GROWTH is a collaborative project among California Institute of Technology (USA), University of Maryland College Park (USA), University of Wisconsin Milwaukee (USA), Texas Tech University (USA), San Diego State University (USA), University of Washington (USA), Los Alamos National Laboratory (USA), Tokyo Institute of Technology (Japan), National Central University (Taiwan), Indian Institute of Astrophysics (India), Indian Institute of Technology Bombay (India), Weizmann Institute of Science (Israel), The Oskar Klein Centre at Stockholm University (Sweden), Humboldt University (Germany), Liverpool John Moores University (UK) and University of Sydney (Australia). 

Based on observations made with the Liverpool Telescope operated on the island of La Palma by Liverpool John Moores University in the Spanish Observatorio del Roque de los Muchachos of the Instituto de Astrofisica de Canarias with financial support from the UK Science and Technology Facilities Council. Based on observations obtained with the Samuel Oschin Telescope 48-inch and the 60-inch Telescope at the Palomar Observatory as part of the Zwicky Transient Facility project. ZTF is supported by the National Science Foundation under Grant No. AST-1440341 and a collaboration including Caltech, IPAC, the Weizmann Institute for Science, the Oskar Klein Center at Stockholm University, the University of Maryland, the University of Washington, Deutsches Elektronen-Synchrotron and Humboldt University, Los Alamos National Laboratories, the TANGO Consortium of Taiwan, the University of Wisconsin at Milwaukee, and Lawrence Berkeley National Laboratories. Operations are conducted by COO, IPAC, and UW. This research used resources of the National Energy Research Scientific Computing Center, a DOE Office of Science User Facility supported by the Office of Science of the U.S. Department of Energy under Contract No. DE-AC02-05CH11231. The 0.7m GROWTH-India Telescope (GIT) is set up by the Indian Institute of Astrophysics (IIA) and the Indian Institute of Technology Bombay (IITB) with support from the Indo-US Science and Technology Forum (IUSSTF) and the Science and Engineering Research Board (SERB) of the Department of Science and Technology (DST), Government of India Grant No.IUSSTF/PIRE Program/GROWTH/2015-16. It is located at the Indian Astronomical Observatory, IIA at Hanle, Ladakh (India). This publication has made use of data collected at Lulin Observatory, partly supported by MoST grant 105-2112-M-008-024-MY3.
The KPED team thanks the National Science Foundation and the National Optical Astronomical Observatory for making the Kitt Peak 2.1-m telescope available. We thank the observatory staff at Kitt Peak for their efforts to assist Robo-AO KP operations.
The KPED team thanks the National Science Foundation, the National Optical Astronomical Observatory, the Caltech Space Innovation Council and the Murty family for support in the building and operation of KPED. SED Machine is based upon work supported by the National Science Foundation under Grant No. 1106171.
The Palomar Gattini-IR project thanks the Mount Cuba Foundation, Heising Simons Foundation, the ANU Futures Scheme, the Binational Science Foundation and Caltech for generous support. 

G.C. Anupama and Varun Bhalerao acknowledge partial support from SERB and IUSSTF.
J.~Sollerman acknowledges support from the Knut and Alice Wallenberg Foundation.
E.~Ofek is grateful for support by  a grant from the Israeli Ministry of Science,  ISF, Minerva, BSF, BSF transformative program, and  the I-CORE Program of the Planning  and Budgeting Committee and The Israel Science Foundation (grant No 1829/12).
P. Gatkine is supported by NASA Earth and Space Science Fellowship (ASTRO18F-0085).
C.-C.~Ngeow, A. Patil and P.-C.~Yu thank the funding from Ministry of Science and Technology (Taiwan) under grants 104-2923-M-008-004-MY5, 106-2112-M-008-007, 107-2119-M-008-012, 107-2119-M-008-014-MY2
E. Bellm and V. Z. Golkhou
acknowledge support from the University of Washington College of Arts and Sciences, Department of Astronomy, and the DIRAC Institute. University of Washington's DIRAC Institute is supported through generous gifts from the Charles and Lisa Simonyi Fund for Arts and Sciences, and the Washington Research Foundation.
E. Bellm
acknowledges support from the Large Synoptic Survey Telescope, which is supported in part by the National Science Foundation through
Cooperative Agreement 1258333 managed by the Association of Universities for Research in Astronomy
(AURA), and the Department of Energy under Contract No. DE-AC02-76SF00515 with the SLAC National
Accelerator Laboratory. Additional LSST funding comes from private donations, grants to universities,
and in-kind support from LSSTC Institutional Members.
E.~Bellm is supported in part by the NSF AAG grant 1812779 and grant \#2018-0908 from the Heising-Simons Foundation.
M.~W.~Coughlin is supported by the David and Ellen Lee Postdoctoral Fellowship at the California Institute of Technology.
Part of this research was carried out at the Jet Propulsion Laboratory, California Institute of Technology, under a contract with the National Aeronautics and Space Administration.
D.~L.~Kaplan was supported by NSF grant AST-1816492.
A.~K.~H.~Kong acknowledges support from the Ministry of Science and Technology of the Republic of China (Taiwan) under grants 106-2628-M-007-005 and 107-2628-M-007-003.
J.\,S.~Bloom and J.\,Martinez-Palomera are partially supported by a Gordon and Betty Moore Foundation Data-Driven Discovery grant. 
Harsh Kumar thanks the LSSTC Data Science Fellowship Program, which is funded by LSSTC, NSF Cybertraining Grant \#1829740, the Brinson Foundation, and the Moore Foundation; his participation in the program has benefited this work.
S. Anand acknowledges support from the PMA Division Medberry Fellowship at the California Institute of Technology.
Rahul Biswas, Ariel Goobar and  Jesper Sollerman acknowledge support from the G.R.E.A.T research environment funded by the Swedish National Science Foundation.
J. Soon acknowledges support by an Australian Government Research Training Program (RTP) Scholarship.

\bibliographystyle{aasjournal}
\bibliography{references}

\end{document}